\documentclass[twocolumn,twocolappendix]{aastex631}
\usepackage{multirow}
\usepackage{color}
\renewcommand\ion[2]{#1$\;${%
\ifx\@currsize\normalsize\small \else
\ifx\@currsize\small\footnotesize \else
\ifx\@currsize\footnotesize\scriptsize \else
\ifx\@currsize\scriptsize\tiny \else
\ifx\@currsize\large\normalsize \else
\ifx\@currsize\Large\large
\fi\fi\fi\fi\fi\fi
\rmfamily{#2}}\relax}

\usepackage{amsmath, amssymb}

\title{SALT3-hostmass}

\newcommand{\UCSC}{Department of Astronomy and Astrophysics, University of California, Santa Cruz, CA 95064, USA}
\newcommand{\JHU}{Department of Physics and Astronomy, The Johns Hopkins University, Baltimore, MD 21218.}
\newcommand{\Einstein}{NASA Einstein Fellow}
\defcitealias{Kenworthy21}{K21}

\newcommand{\nohost}{{\tt SALT3.HostIgnore}}
\newcommand{\hostnew}{{\tt SALT3.HostResid}}

\begin{document}

\title{A Spectroscopic Model of the Type Ia Supernova--Host-Galaxy Mass Correlation from SALT3}

\author[0000-0002-6230-0151]{D.~O.~Jones}
\altaffiliation{\Einstein}
\affiliation{\UCSC}

\author[0000-0002-5153-5983]{W.~D.~Kenworthy}
\affiliation{\JHU}
\affiliation{The Oskar Klein Centre for Cosmoparticle Physics, Department of Physics,Stockholm University, SE-10691 Stockholm, Sweden.}
\author[0000-0002-5995-9692]{M.~Dai}
\affiliation{\JHU}

\author[0000-0002-2445-5275]{R.~J.~Foley}
\affil{\UCSC}

\author[0000-0003-3221-0419]{R.~Kessler}
\affil{Kavli Institute for Cosmological Physics, University of Chicago, Chicago, IL 60637, USA} 
\affil{Department of Astronomy and Astrophysics, University of Chicago, 5640 South Ellis Avenue, Chicago, IL 60637, USA}

\author[0000-0002-2361-7201]{J.~D.~R.~Pierel}
\affil{Space Telescope Science Institute, Baltimore, MD 21218, USA}

\author[0000-0003-2445-3891]{M.~R.~Siebert}
\affil{\UCSC}


\begin{abstract}
  
The unknown cause of the correlation between Type Ia supernova (SN\,Ia) Hubble residuals and their host-galaxy masses (the ``mass step") may bias cosmological parameter measurements.  To better understand the mass step, we develop a SALT3 light-curve model for SN cosmology that uses the host-galaxy masses of 296 low-redshift SNe\,Ia to derive a spectral-energy distribution--host-galaxy mass relationship.  The resulting model has larger \ion{Ca}{II} H\&K, \ion{Ca}{II} near-infrared triplet, and \ion{Si}{II} equivalent widths for SNe in low-mass host galaxies at 2.2-2.7$\sigma$ significance; this indicates higher explosion energies per unit mass in low-mass-hosted SNe.  The model has phase-dependent changes in SN\,Ia colors as a function of host mass, indicating intrinsic differences in mean broadband light curves. Although the model provides a better fit to the SN data overall, it does not substantially reduce data--model residuals for a typical light curve in our sample nor does it significantly reduce Hubble residual dispersion.  This is because we find that previous SALT models parameterized most host-galaxy dependencies with their first principal component, although they failed to model some significant spectral variations.  Our new model is luminosity and cosmology independent, and applying it to data reduces the mass step by $0.021\pm0.002$~mag (uncertainty accounts for correlated data sets); these results indicate that $\sim$35\% of the mass step can be attributed to luminosity-independent effects.  This SALT model version could be trained using alternative host-galaxy properties and at different redshifts, and therefore will be a tool for understanding redshift-dependent correlations between SNe\,Ia and their host properties as well as their impact on cosmological parameter measurements.


\end{abstract}

\section{Introduction}

Type Ia supernovae (SNe\,Ia) are reliable distance indicators across more than 10~Gyr of cosmic history \citep{Jones13,Rubin13,Rodney14}, and are a mature cosmological tool for measurements of the Hubble constant \citep{Hamuy21,Riess21b} and the dark energy equation of state, $w$ \citep{Scolnic18,Jones19,Abbott19,Brout22}.  However, a small unexplained $\sim$0.06-mag shift in their distance measurements as a function of their host-galaxy masses \citep{Kelly10,Lampeitl10,Sullivan10} could be an indication of systematic uncertainties.

In addition to the host-galaxy mass ``step", numerous other correlations between SNe\,Ia and their host-galaxy properties have been found in the years since its discovery.  These include potential correlations with the host-galaxy metallicity \citep{Hayden13,Rose21}, the host-galaxy star formation rate \citep{Rigault13}, and the host-galaxy stellar age \citep{Childress14,Rose19}.  Other correlations have been seen with the properties of the host galaxy near the SN location, including the local star formation rate or specific star formation rate \citep{Rigault13,Kim18,Rigault20}, the local stellar mass \citep{Jones18}, and the local $U-V$ or $U-R$ color \citep{Roman18,Kelsey21,Kelsey22}.  It is not only SN\,Ia inferred distance measurements that correlate with host properties, but also SN shape and color \citep[e.g.,][]{Childress13}, and the SN\,Ia color--luminosity relation \citep[e.g.,][]{GonzalezGaitan21}.

The reasons for these numerous observed correlations are unclear, and are complicated by the inherent correlations between the galaxy properties themselves.  Similarly, SN properties such as stretch and color are correlated with galaxy properties \citep{Hamuy95,Branch96,Hamuy00, Howell01,Gallagher05}, and therefore model-dependent corrections for, for example, the stretch--luminosity relation \citep{Phillips93,Tripp98}, have implicit host-galaxy dependencies.  Proposed explanations for the mass step and related effects include a possible change in the properties of the SNe\,Ia due to progenitor metallicities \citep[e.g.,][]{Rose21}, ``prompt" versus ``delayed" progenitor ages  \citep[e.g.,][]{Childress14}, or variations in the extinction law of extrinsic dust \citep{Brout21,Popovic21b}.

This final hypothesis gives a prediction that the appearance of a step in SN\,Ia Hubble residuals should be strongly wavelength dependent due to the higher dust attenuation at bluer wavelengths.  \citet{Ponder21} and \citet{Uddin20} therefore looked for evidence of the host-galaxy mass step at near-infrared (NIR) wavelengths, as this would disfavor an extinction-based cause of the mass step, and recovered mass steps in the $H$-band with 2--3$\sigma$ significance.  \citet{Thorp21} used the hierarchical Bayesian model BayeSN \citep{Mandel22} to constrain the distribution of the total-to-selective dust extinction ratio, $R_V$ in low- and high-mass galaxies, finding a statistically insignificant difference between the two populations.  \citet{Jones22} also found $\sim$2-$\sigma$ evidence for the mass step using both low- and high-redshift SN\,Ia data observed in the rest-frame NIR. 

On the other hand, \citet{Johansson21} found a mass step that decreased in the NIR (particularly the $JH$ bands), and that disappeared when $R_V$ was included as a free parameter in the SN\,Ia distance fitting, although it is possible that fitting for $R_V$ could have added noise to the distances.  Additional evidence for scatter in redder SNe\,Ia was seen by \citet{Rose22} and evidence for correlations between host-galaxy $R_V$ variation and Hubble residuals was seen recently by \citet{Meldorf22,Kelsey22,Wiseman22}, though in these studies either a small mass step or a host-galaxy color-dependent step remained after modeling the host mass--host $R_V$ relation.

Exploring the wavelength dependence of the mass step has therefore proven to be a powerful tool for understanding its underlying physical mechanisms.  A spectroscopically resolved model of the dependence of the color-corrected SN\,Ia spectral energy distribution (SED) on host-galaxy mass could be an even more precise indicator for distinguishing between different possible underlying physical effects.  This was first attempted in \citet{Siebert20} by using a set of composite spectra generated from the Kaepora model \citep{Siebert19} and explored as a systematic test in \citet{Pierel21}, but the composite spectra method is unable to fully control for the differences between SN shape and color in different galaxy populations because it uses averaging and approximate binning to sample highly non-Gaussian distributions.  In addition, recent SN\,Ia distance determination methods, such as \citet{Boone21a,Boone21}, have seen evidence that improved SN\,Ia modeling techniques reduce the size of the mass step.

\subsection{A New Approach for Building a SN\,Ia--Host-Galaxy Mass Model}

Fundamentally, each of these host-galaxy step measurements have similar methodologies: they generate Hubble residuals using a model that does not include host-galaxy properties and then subsequently explore the ways in which those Hubble residuals correlate with host properties.  Recent work to build the SALT3 model for SN\,Ia standardization \citep[hereafter \citetalias{Kenworthy21}]{Kenworthy21} has created a new opportunity to study the fine-grained correlations between SN\,Ia spectra with their host galaxies by incorporating host properties in the generation of the model itself.

To enable this work, \citetalias{Kenworthy21} created an open-source training code called {\tt SALTshaker} and added several improvements to the SALT2 model framework \citep{Guy07,Guy10,Betoule14,Taylor21} that has been used to measure SN\,Ia light-curve parameters for nearly all measurements of dark energy properties in the past decade.  {\tt SALTshaker} constructs a SALT3 model by finding flux surfaces as a function of phase and wavelength that are associated with the zeroth and first principal components of SN\,Ia variation, where the zeroth component is a mean SN\,Ia SED and the first component correlates strongly with the SN\,Ia light-curve stretch.  Simultaneously, it infers a phase-independent color law to model the effects of reddening. \citet{Dai23} validated the model training framework, \citet{Taylor23} tested the model's performance against SALT2, and the model was extended to the NIR in \citet{Pierel22}.

Here, we retrain the SALT3 model after adding a ``host-galaxy mass" flux component as a function of phase and wavelength to the model framework.  We train two models: first, we train nominal SALT3 model surfaces and subsequently add a host-galaxy component while keeping the baseline parameters and surfaces fixed; in this case the host-galaxy component is a model of the ``missing" features from our current SN\,Ia standardization paradigm that vary with host-galaxy mass. Second, we train all model surfaces simultaneously, and the host-galaxy component is defined to be the difference between the mean SN\,Ia SED in a high- versus low-mass galaxy.  The goal of this work is to understand the properties of the SNe\,Ia that could drive the host-galaxy mass step, and develop a framework for building a wavelength- and phase-dependent understanding of the ways in which the SN--host connection could bias cosmological parameter measurements.

In Section \ref{sec:model} we describe the revised model framework, including alterations to the training sample and our host-galaxy mass measurements.  In Section \ref{sec:results} we present the trained model surfaces as well as the resulting Hubble residuals and mass-step measurements.  In Section \ref{sec:discussion} we discuss future steps, and in Section \ref{sec:conclusions} we conclude.

\section{Building the SALT3 Host-Galaxy Model}
\label{sec:model}

\begin{table*}[]
    \centering
    \caption{The SALT3 Host-Galaxy Training Sample ($z < 0.15$)}
    \begin{tabular}{lrrrrrrr}
    \hline \hline\\[-1.5ex]
    &\multicolumn{2}{c}{${\rm log(M_{\ast}/M_{\odot})} < 10$}&&\multicolumn{2}{c}{${\rm log(M_{\ast}/M_{\odot})} > 10$}&&\\
    \cline{2-3} \cline{5-6}
    Survey & N$_{\mathrm{SN}}$ &N$_{\mathrm{spectra}}$ && N$_{\mathrm{SN}}$ &N$_{\mathrm{spectra}}$&Filters&Reference\\
    \hline\\[-1.5ex]
Calan-Tololo&1&0&&4&0&$BV\!RI$&\citet{Hamuy96}\\
CfA1&1&7&&7&59&$U\!BV\!RI$&\citet{Riess99}\\
CfA2&4&70&&9&96&$U\!BV\!RI$&\citet{Jha06}\\
CfA3&12&135&&39&399&$U\!BV\!RIri$&\citet{Hicken09a}\\
CfA4&9&0&&21&0&$BV\!ri$&\citet{Hicken12}\\
CSP&3&5&&10&31&$uBV\!gri$&\citet{Krisciunas17}\\
Foundation&70&52&&81&60&$griz$&\citet{Foley18}\\
Misc. low-$z$&2&9&&23&207&$U\!BV\!RI$&\citet{Jha07}\\
\hline\\*[-1.5ex]
Total&102&278&&194&852&\nodata&\nodata\\
\hline\\*[-1.5ex]
    \end{tabular}
    \label{table:sample}
\end{table*}

The baseline SALT3 framework creates a model of SNe\,Ia using three parameters: the amplitude of the SN, $x_0$, a parameter describing first-order variability, $x_1$, and the phase-independent color, $c$.  The parameter $x_1$ is often referred to as a ``stretch" parameter, but it also encodes other wavelength- and phase-dependent intrinsic variability beyond a simple stretching of the light curve.  These parameters, combined with the trained principal-component-like model surfaces $M_0(p,\lambda)$ and $M_1(p,\lambda)$, and the color law $CL(\lambda)$, are used to yield a model of the phase ($p$)- and wavelength ($\lambda$)- dependent flux of a given SN $F(p,\lambda)$:

\begin{align}
    F(p,\lambda) = &x_0 [M_0(p,\lambda) + x_1 M_1(p,\lambda)] \nonumber\\ & \cdot \exp(c \cdot CL(\lambda)).
\label{eqn:salt3}
\end{align}

\noindent This model framework has proven successful at standardizing SNe\,Ia to distance precision of up to $\sim$5\%, and has several advantages over other SN\,Ia standardization models, including the following:

\begin{enumerate}
    \item The SALT3 model surfaces are defined using third-order $B$-spline bases with resolution of 72\AA, allowing the SALT3 model to include native $K$-corrections.
    \item The SALT3 model includes the SN amplitude $x_0$ as a free parameter, which allows the use of high-$z$ data to provide a well-calibrated, rest-frame ultraviolet (UV) training set without making the model dependent on cosmological parameters.
    \item The open-source, {\tt Python}-based {\tt SALTshaker} training code can be developed and improved as new SN\,Ia samples are built from, for example, the Zwicky Transient Facility \citep{Bellm19}, the Young Supernova Experiment \citep{Jones21}, the Rubin Observatory \citep{DESC18}, and the {\it Roman Space Telescope} \citep{Rose21}.
\end{enumerate}

More details on the model and a discussion of improvements compared to the widely used SALT2 model \citep{Guy10} are given in \citetalias{Kenworthy21}.

The SALT3 training procedure and data are also described in detail in \citetalias{Kenworthy21}.  In brief, both photometric and spectroscopic training data are compared to the initial model surfaces, which are adjusted iteratively to minimize the $\chi^2$ with a Levenberg--Marquardt algorithm.  Regularization terms penalize the $\chi^2$ for high-frequency variations in the model surfaces, effectively smoothing those surfaces in regions without sufficient data.  Constraints are used to remove degeneracies between the model components ($M_0$, $M_1$, $CL$) and parameters ($x_0$, $x_1$, $c$).  For example, the mean and standard deviation of $x_1$ are set to $(\mu_{x_1},\sigma_{x_1})=(0,1)$ across the sample to avoid a degeneracy in which a larger spread in $x_1$ and a lower-amplitude $M_1$ gives the same model fluxes as a smaller $x_1$ spread and a higher-amplitude $M_1$.

There are two uncertainty components in the SALT3 model:  the ``in-sample" model uncertainties are used to model the component of SN\,Ia intrinsic variation that is not included in the SALT3 model framework, and the ``out-of-sample" model uncertainties are due to having a finite training sample size.  The in-sample uncertainties are iteratively estimated during the training process using a log-likelihood approach, while the out-of-sample uncertainties are estimated from an inverse Hessian matrix after the training process has concluded.

The SALT3 training process was validated on SN simulations in \citetalias{Kenworthy21}, with subsequent tests performed by \citet{Pierel22}, \citet{Dai23}, and \citet{Taylor23}.

\subsection{The SALT3 Host Galaxy Mass Model}

\begin{table*}[]
    \centering
    \caption{Host-Galaxy Photometric Data}
    \begin{tabular}{lrrr}
        Instrument & Filters & PSF FWHM (\arcsec) & Reference\\
        \hline \hline
         GALEX & $NUV$, $FUV$ &4.5-5.4 & \citet{Martin05}\\
         SDSS &$ugriz$& 1.3-1.5 & \citet{York00}\\
         PS1 & $grizy$& 1.02-1.31 & \citet{Flewelling20}\\
         2MASS &$JHK$& 2.8-2.9 & \citet{Skrutskie06}\\
         \hline
    \end{tabular}
    \label{table:hostphot}
\end{table*}

Here, we build a new version of SALT3 that includes an explicit dependence on host-galaxy mass.  To include host-galaxy properties in the model framework, we add a host-galaxy parameter, $x_{host}$, and model surface, $M_{host}$:

\begin{align}
  \label{eqn:salt3host}
    F(p,\lambda) = &x_0 [M_0(p,\lambda) + x_1 M_1(p,\lambda) + \\ \nonumber
    &x_{host} M_{host}(p,\lambda)] \cdot \exp(c \cdot CL(\lambda)).
\end{align}

\noindent In the current analysis, the host-galaxy parameter is simply chosen to be $+1/2$ for a SN in a host galaxy with ${\rm log(M_{\ast}/M_{\odot})} > 10$ and $-1/2$ for a SN in a host with ${\rm log(M_{\ast}/M_{\odot})} < 10$.  This choice allows the $M_{host}$ surface to be interpretable as the difference between a mean high-mass-hosted SN versus a mean low-mass-hosted SN.  We use the canonical location of the mass step of ${\rm log(M_{\ast}/M_{\odot})} = 10$ (e.g., \citealp{Sullivan10} and subsequent studies), even though the median host-galaxy mass from the data presented in Section \ref{sec:trainingdata} is approximately 10.3~dex; this choice reduces the signal-to-noise ratio (S/N) of our resulting mass-step component, but it ensures that our results will have direct relevance to the mass step measured in cosmological analyses.

We found that the uncertainties on the host-galaxy masses are sufficiently small that the effect of including them would be negligible, but a future framework that includes the uncertainty on host-galaxy parameters could be necessary for incorporating additional host-galaxy properties into the model.

We use this formalism to train a SALT3 model that includes host-galaxy mass information in two ways, described below:

\begin{enumerate}
    \item We use our low-$z$ training data (Section \ref{sec:trainingdata}) to determine $M_0$, $M_1$ and $CL$ assuming the original SALT model formalism (Equation \ref{eqn:salt3}).  Then, while keeping $M_0$, $M_1$, and the color law fixed, we add the host component to the SALT model formalism and train again.  The derived $M_{host}$ component contains only spectroscopic and photometric features that depend on host mass and are ``missing" from existing SALT models.  We refer to this model as \hostnew.
    \item We train the full model, allowing $M_0$, $M_1$, and $CL$ to vary simultaneously with $M_{host}$.  In this case, the derived $M_{host}$ component is the change in the mean spectrum of a SN\,Ia between low- and high-mass host galaxies.  Additional host-independent variability is captured by a redefined $x_1$ parameter and $M_1$ surface.  We refer to this model as {\tt SALT3.Host}.
\end{enumerate}

\noindent These models are compared to the \nohost\ model, which is a model trained using the traditional SALT formalism (Equation \ref{eqn:salt3}) with the exact same training data as {\tt SALT3.Host} and \hostnew.

For the {\tt SALT3.Host} model, to separate the effects of host-galaxy mass from the component of intrinsic variability that is independent of the host properties, we include a constraint to set the correlation between $x_1$ and $x_{host}$ to zero.\footnote{This includes both a prior in the SALT3 training itself and updated definitions after the training has concluded: $x_1^{\prime} = x_1 + \alpha x_{host}$ and $M_{host}^{\prime} = M_{host} - \alpha M_1$.  The variable $\alpha$ is a constant that is scaled to remove the $x_1$,$x_{host}$ correlation.}
This is similar to the existing SALT3 constraint that the correlation between $x_1$ and $c$ is zero in the training sample; the $x_1$,$c$ constraint has the physically intuitive meaning that if the color is related to dust extinction, it should not depend on the stretch. 


This $x_1$,$x_{host}$ constraint redefines the $x_1$ parameter as the phase-dependent SN variability that is {\it uncorrelated} with the host-galaxy mass. Correspondingly, $M_{host}$ can be interpreted as the difference in the mean host-galaxy spectrum between a SN\,Ia in a high-mass versus a low-mass host galaxy.

As the SALT3 training is cosmology independent, due to including no prior on the amplitude parameter $x_0$, the overall amplitude of $M_{host}$ is degenerate with $x_0$.  Therefore, we add offsets to both $M_1$ and $M_{host}$ such that they have zero flux when integrated over the $B$ bandpass at maximum light.  We note that without this definition the size of the measured mass step from this model would be directly correlated with the $B$-band luminosity of the mass component; the $M_{host}$ luminosity can be defined such that it removes the mass step entirely for a given data set, but the $M_{host}$ luminosity cannot be determined by {\tt SALTshaker} itself due to the degeneracy between $x_{host}$ and $x_0$ for each SN.

Finally, the nominal SALT3 error model includes the in-sample and out-of-sample variances for $M_0$ and $M_1$ as well as the diagonal covariance between $M_0$ and $M_1$.  We add to this model the variance of $M_{\rm host}$ and the diagonal covariance between $M_0$ and $M_{\rm host}$ ($M_{host}$ and $M_1$ are defined as described above to have negligible covariance with each other).  Because we found some evidence that {\tt SALTshaker}  overestimates the model uncertainties when trained on a small sample, perhaps due to its regularization assumptions, we used a simplified approach for uncertainty computation for the out-of-sample uncertainties in this model: we bootstrap resampled the input SNe 50 times and retrained the model on each new sample after reducing the regularization terms by a factor of 100.

\subsection{Training Data}
\label{sec:trainingdata}

To build the \hostnew\ and {\tt SALT3.Host} models, we restrict to training data at redshifts $z < 0.15$ to ensure that host-galaxy masses can be measured robustly.  This allows us to include Galaxy Evolution Explorer ({\it GALEX}) and Two Micron All Sky Survey (2MASS) photometry for the SN host galaxies, which are not deep enough to detect many higher-redshift galaxies, and it avoids potential concerns of biases in high-$z$ mass estimates \citep[e.g.,][]{Paulino-Alfonso22}.  This choice adds significant noise to the model training below $\sim$3500\AA\ due to the lack of well-calibrated near-UV data from high-redshift surveys --- and perhaps avoids systematic redshift-dependent differences at those wavelengths \citep{Foley12} --- but allows more robust mass estimates.

We also update the photometry of the training data relative to \citetalias{Kenworthy21} to use the ``SuperCal-Fragilistic" cross-calibration update to these photometry presented in \citet{Brout21b}; this model is published in \citet{Taylor23}.  SuperCal-Fragilistic uses Pan-STARRS observations of tertiary standard stars employed by different SN surveys, along with updates to the {\it Hubble Space Telescope} absolute calibration from \citet{Bohlin20}, to put individual SN surveys on a common photometric system.  We also update the Carnegie Supernova Project (CSP) photometry to use the third CSP data release \citep[DR3;][]{Krisciunas17}; \citetalias{Kenworthy21} used CSP Data Release 2 to match the cross-calibration analysis done by \citet{Scolnic15}, whereas \citet{Brout21b} uses DR3.  A full SALT3 model using the recalibrated \citetalias{Kenworthy21} data is publicly available as part of the {\tt sncosmo} package \citep{sncosmo} as well as SNANA \citep{Kessler10}.

A summary of the {\tt SALT3.Host} training data is given in Table \ref{table:sample}.  We do not attempt to extend this model to the NIR \citep[e.g.,][]{Pierel22} because one of our primary goals is to understand spectral variation and there are currently just 51 publicly available NIR spectra of SNe\,Ia that have well-sampled NIR light curves.

\begin{figure}
    \centering
    \includegraphics[width=3.5in]{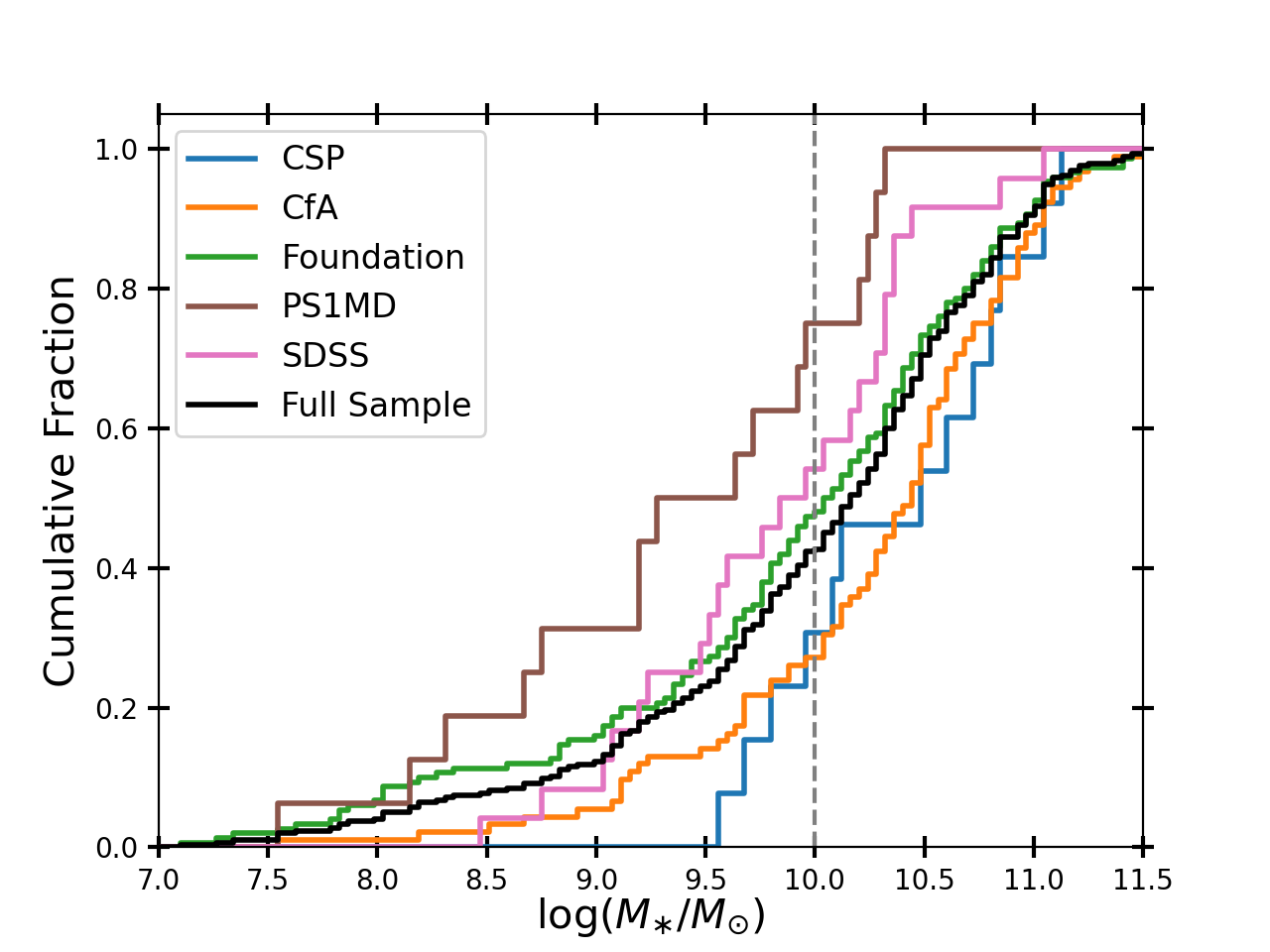}
    \caption{Cumulative distributions of host-galaxy masses for the different surveys included in the training sample.  Approximately 75\% of our masses are from the Foundation and CfA samples.}
    \label{fig:hostmass}
\end{figure}

\begin{figure}
    \centering
    \includegraphics[width=3.4in]{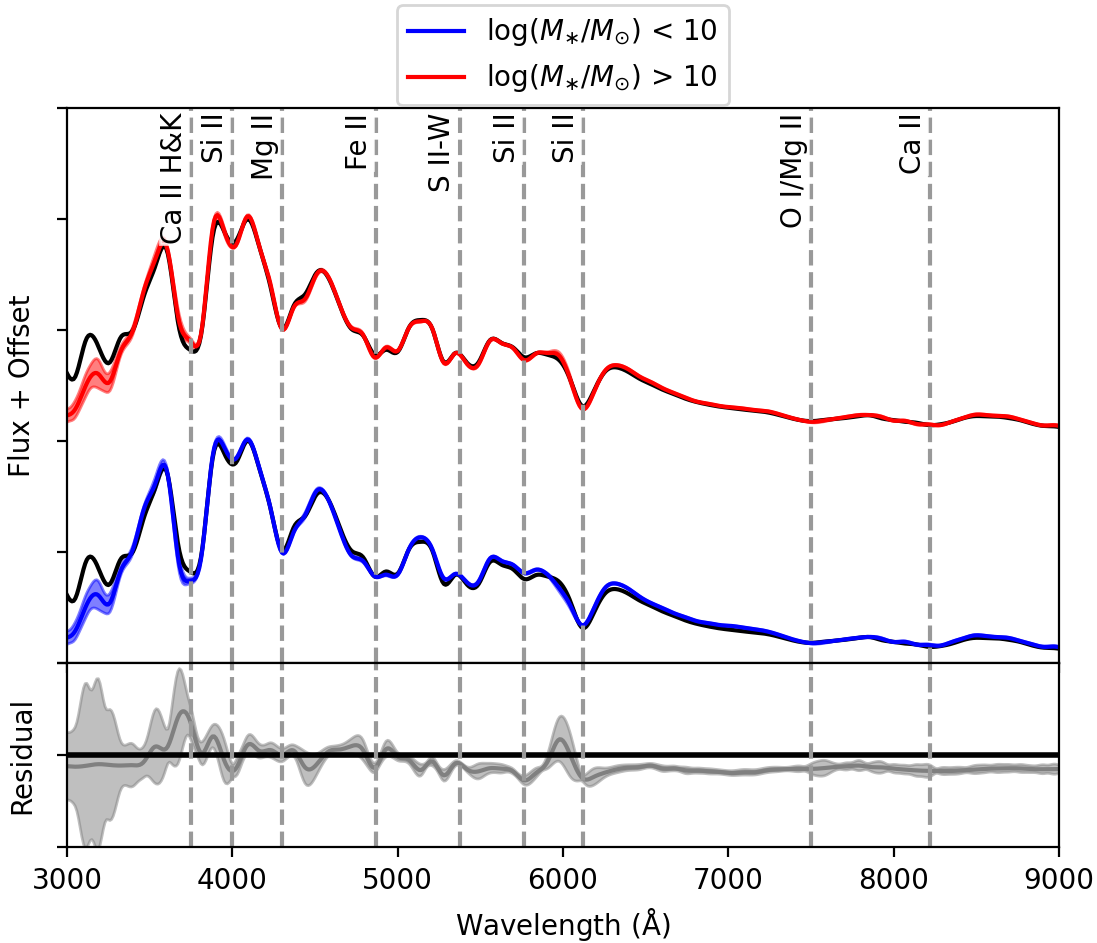}
    \caption{Top: the average spectrum of a $(x_1,c) = (0,0)$ SN\,Ia at maximum light for SNe in high-mass (red) and low-mass (blue) host galaxies. The $(x_1,c) = (0,0)$ spectrum from \citetalias{Kenworthy21} is in black.  Bottom: the difference between the red and blue spectra in the top panel, equal to the $M_{host}$ component from Equation \ref{eqn:salt3host}.  We note that both models differ from \citetalias{Kenworthy21} in the blue due to the removal of high-redshift, rest-frame UV data.}
    \label{fig:peak_spec}
\end{figure}

\begin{figure*}
    \centering
    \includegraphics[width=7in]{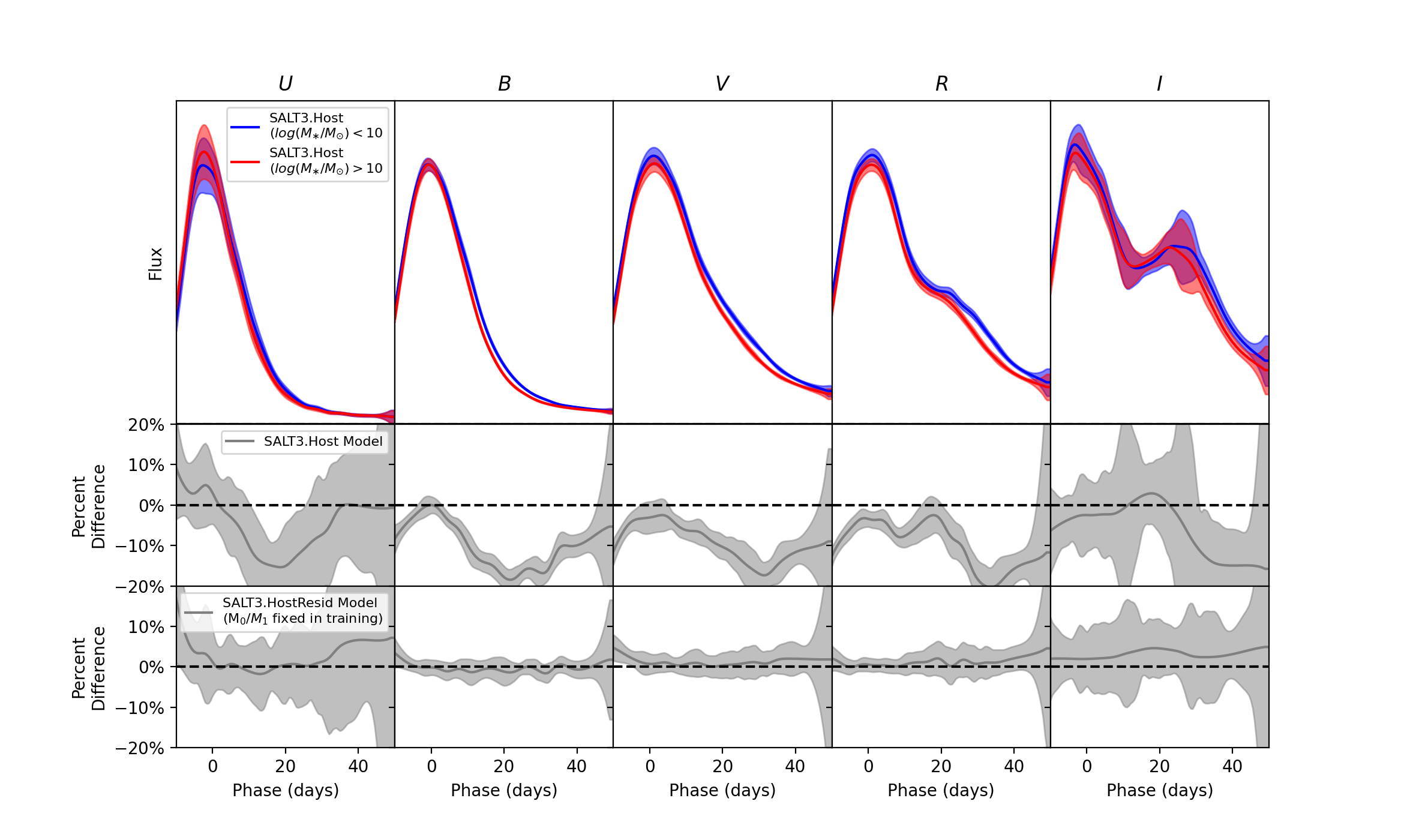}
    \caption{Top: $U\!BV\!RI$ light curves of the mean SN\,Ia in a high-mass (red) vs.\ low-mass (blue) host galaxy from the {\tt SALT3.Host} model ($M_0 \pm 0.5M_{host}$).  The $y$-axis is scaled from zero to $+$130\% of the flux at maximum light in the high-mass light curve for each band; note that red and blue curves are normalized by {\tt SALTshaker} to have the same flux at $B$-band maximum light.  
    Middle and bottom: residual curves for the difference between SNe in high- and low-mass hosts for the {\tt SALT3.Host} model (middle) and the \hostnew\ model (bottom).}
    \label{fig:lc}
\end{figure*}

\begin{figure}
    \centering
    \includegraphics[width=3.4in]{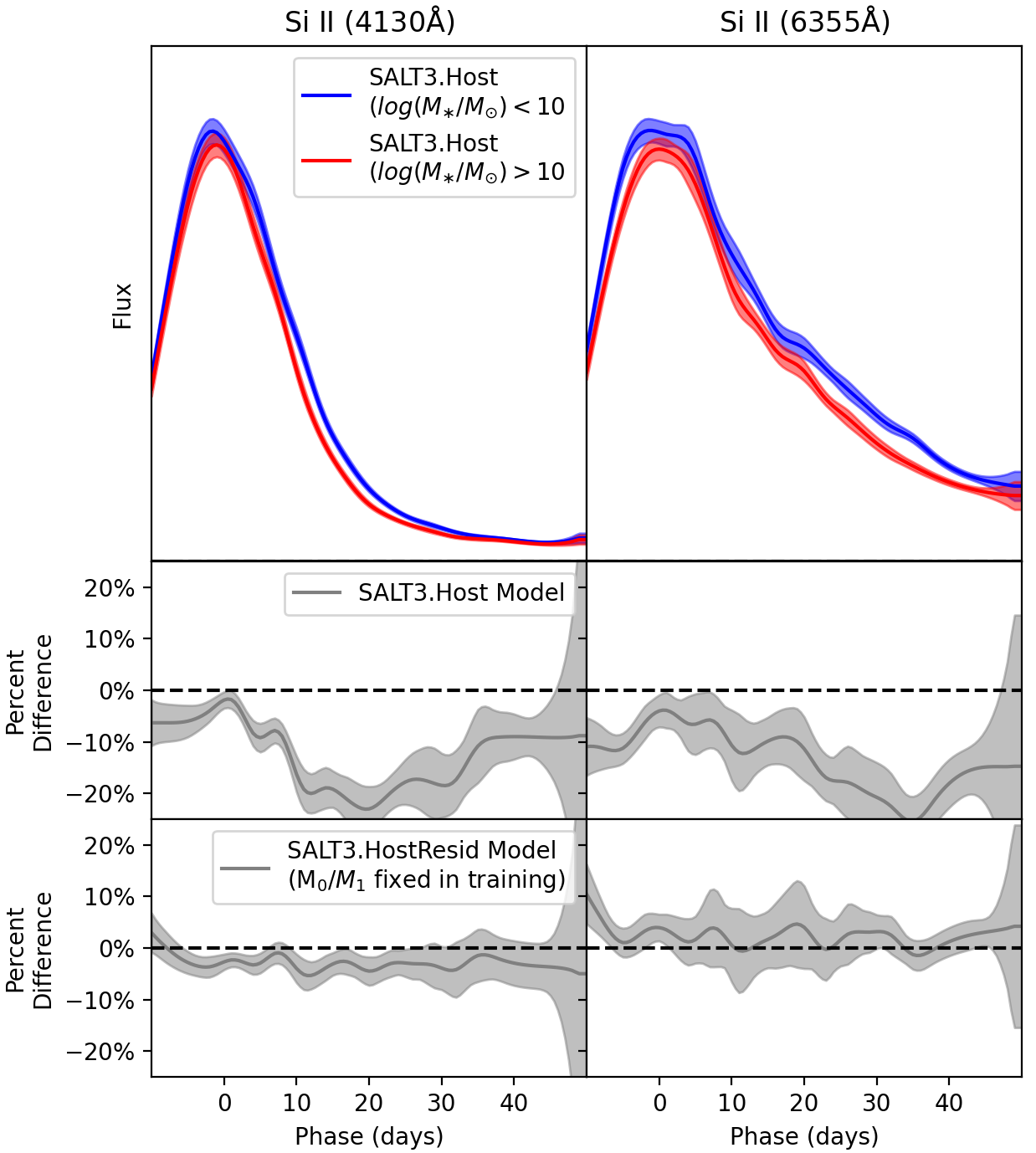}
    \caption{Same as Figure \ref{fig:lc} but for top-hat filters centered on the \ion{Si}{II} lines at rest-frame 4130 and 6355\AA.  To cover the full lines while avoiding nearby absorption features, we use filters from 3900 to 4100~\AA\ and 6000-6250~\AA, respectively (\ion{Si}{II} is blueshifted from its rest frame in SNe\,Ia).  We note that there is greater noise in the line-profile light curves than in the broadband light curves in Figure \ref{fig:lc}.}
    \label{fig:lc_si}
\end{figure}

\subsection{Measuring Host-Galaxy Masses}

We measure host-galaxy masses for SNe in our training sample by first using the Galaxies HOsting Supernova Transients \citep[GHOST;][]{Gagliano21} software to determine the host galaxy for each SN.  GHOST uses a gradient-ascent algorithm to match SN locations to their likely host galaxies using deep PS1 postage-stamp images at each SN location.  Next, we visually inspect the GHOST matches, and alternatively use the SExtractor-based ``directional light radius" (DLR) method \citep{Sullivan06,Gupta16} on $r$-band images from Pan-STARRS  \citep[PS1;][]{Chambers16} to correct the few percent of erroneous matches.  Finally, for SNe with no identifiable host either visually or with GHOST, we assume the true host is at the SN location; at $z < 0.15$, hosts that are undetected in PS1 imaging will all have ${\rm log(M_{\ast}/M_{\odot})} < 10$, but we use their location for photometric measurements in case upper limits on the mass are needed for future work.

With these host-galaxy locations, we use SExtractor \citep{Bertin96} to measure the isophotal elliptical radius of each galaxy, again using PS1 $r$-band imaging.  We measure photometry within this radius using images from {\it GALEX}, the Sloan Digital Sky Survey (SDSS), PS1, and 2MASS, modifying the radius slightly to account for the changing size of the point-spread function (PSF) for each filter and instrument (Table \ref{table:hostphot}).

Finally, we use LePHARE \citep{Ilbert06} to measure the host-galaxy masses of each SN.  We used the \citet{Chabrier03} initial mass function and the \citet{Bruzual03} SED template library.  The extinction $E(B-V)$ is varied from $0$-$0.4$~mag.  The resulting host-galaxy mass distributions for each subsample in our analysis are shown in Figure \ref{fig:hostmass}.

\section{Results}
\label{sec:results}

The {\tt SALT3.Host} model is illustrated in Figure \ref{fig:peak_spec} for spectra at maximum light and in Figure \ref{fig:lc} for $U\!BV\!RI$ light curves.    The phase-dependent evolution of the \ion{Si}{II} lines are shown in Figure \ref{fig:lc_si} and the (small) differences in color laws between models are shown in Figure \ref{fig:colorlaws}.  The full $M_{host}$ spectral component is shown in Figure \ref{fig:spec_sequence} for the {\tt SALT3.Host} model and in Figure \ref{fig:spec_sequence_fixed} for the \hostnew\ model.  Finally, the phase-dependent $B-V$ colors of the {\tt SALT3.Host} model are shown in Figure \ref{fig:color}.

Below, we discuss the characteristics of these surfaces.  We note that uncertainties in the light curves are well correlated on scales of $\sim$3~days, but are largely independent over longer timescales.  In addition, the figures and analysis in this section use only the out-of-sample (bootstrapped) uncertainties, as these are the formal statistical uncertainties on the determination of the model surfaces.  We include in-sample variance in the distance fitting (Section \ref{sec:distances}).

\subsection{Model Spectra and Light Curves}

We first examine the {\tt SALT3.Host} model, and find that spectra at maximum light have significantly different \ion{Ca}{H\&K} and \ion{Si}{II} (6355\AA) line profiles as a function of host-galaxy mass.  In particular, SNe in low-mass hosts have broader absorption features on average (Figure \ref{fig:peak_spec});
at maximum light, we measure larger Ca NIR triplet, \ion{Ca}{H\&K} and \ion{Si}{II} equivalent widths at 2.7-, 2.3-, and 2.2-$\sigma$ significance, respectively, which are correlated with higher velocities and explosion energies per unit of SN mass \citep[e.g.,][]{Wang09,Zhao15}\footnote{We note that the blue component of \ion{Ca}{H\&K} may be contaminated by \ion{Si}{II} at 3858\AA\ \citep{Foley13lines}, but the difference between these features still indicates host-mass-dependent absorption due to intermediate-mass elements.}.    The \ion{Ca}{H\&K} velocity is higher in low-mass hosts, but at just 1.3-$\sigma$ significance, and there is no difference in the \ion{Si}{II}
velocity\footnote{Line velocities and equivalent widths are measured following \citet{Siebert20}.  Errors are from 50 Gaussian-random realizations of each spectrum, where the Gaussian standard deviation corresponds to the wavelength-dependent noise.}.  While the significance of these results could be somewhat enhanced by the look-elsewhere effect, we note that previous studies have specifically investigated differences in \ion{Si}{II} and \ion{Ca}{H\&K} absorption features as potentially correlated with Hubble residuals \citep{Foley11,Siebert20}, so we have {\it a priori} reasons to suspect host-dependent changes in these line profiles.  
    
Additional spectral differences are observed at later phases, particularly in the \ion{Si}{II} features (Figures \ref{fig:lc_si} and \ref{fig:spec_sequence}). The differences in light-curve properties are not limited to places where spectral features are expected to vary (e.g., we observe qualitatively similar behavior at $\sim$4500-5000\AA), but variation in these \ion{Si}{II} and other absorption features could explain some of the variation present in the broadband light curves. We also see other differences between the high-mass host galaxy and low-mass host galaxy SN models at 25--35 days (Figures \ref{fig:lc} and \ref{fig:spec_sequence}), near the time of NIR maximum light, which is an epoch that could be particularly sensitive to the metallicity of the SN progenitor \citep{Kasen06}.

In Figure \ref{fig:lc}, we also confirm previous results that SNe\,Ia in high-mass hosts have narrower shapes on average than those in low-mass hosts \citep[e.g.,][]{Hamuy00,Gallagher05}, as well as an earlier time of secondary maximum light (Figure \ref{fig:lc}).  Significant phase-dependent color differences as a function of host mass can be seen (Figure \ref{fig:color}), which were previously captured as part of the SALT3 $M_1$ component and represented by the SALT3 $x_1$ component in light-curve fits.

Lastly, we note that the updated $M_1$ component appears qualitatively similar to the \nohost\ model, but with larger-amplitude $x_1$-dependent color changes, particularly near maximum light (these are phase-dependent colors and can therefore be separated from the phase-independent color law).

We compare these {\tt SALT3.Host} results to light-curve residuals and spectra from the \hostnew\ model in Figures \ref{fig:lc} and \ref{fig:lc_si} (bottom panels) and Figure \ref{fig:spec_sequence_fixed}.  Near maximum light, we see that previous SALT models have not fully accounted for variations in the \ion{Ca}{H\&K} and \ion{Si}{II} absorption features between different host-galaxy types.
There is also a slight possibility of increased amplitude of the host-mass component in the $I$ band, indicated by the persistent low-significance offset in the bottom-right panel of Figure \ref{fig:lc}.  Overall, however, the pre-existing $M_1$ surface had encoded most of the host-dependent variation in SN properties, and the amplitudes of the spectral changes in Figure \ref{fig:spec_sequence_fixed} are significantly smaller than those in Figure \ref{fig:spec_sequence}.

\subsubsection{Colors}
\label{sec:color}

In Figure \ref{fig:color}, we show the evolution of $B-V$ synthetic color for the {\tt SALT3.Host} model.  We see significantly different color evolution from $\sim$10 to 20 days after maximum light; the evolving color difference may indicate that these colors are intrinsic to the SNe. We note that qualitatively similar behavior is seen in Figure 11 of \citet{Siebert20}, when comparing SNe with negative (preferentially high-mass hosts) versus positive (preferentially low-mass hosts) Hubble residuals.

The \hostnew\ model, on the other hand, has a statistically insignificant color shift (Figure \ref{fig:lc}), implying that adding a host-galaxy mass component to the SALT model has a negligible effect on the fitted broadband light curves of a given SN. We do note that if the persistent low-significance color shift in the $I$ band shown in Figure \ref{fig:lc} is both real and due to extrinsic dust, it may imply lower $R_V$ for high-mass galaxies as predicted by \citet{Brout21}.

\begin{figure}
    \centering
    \includegraphics[width=3.5in]{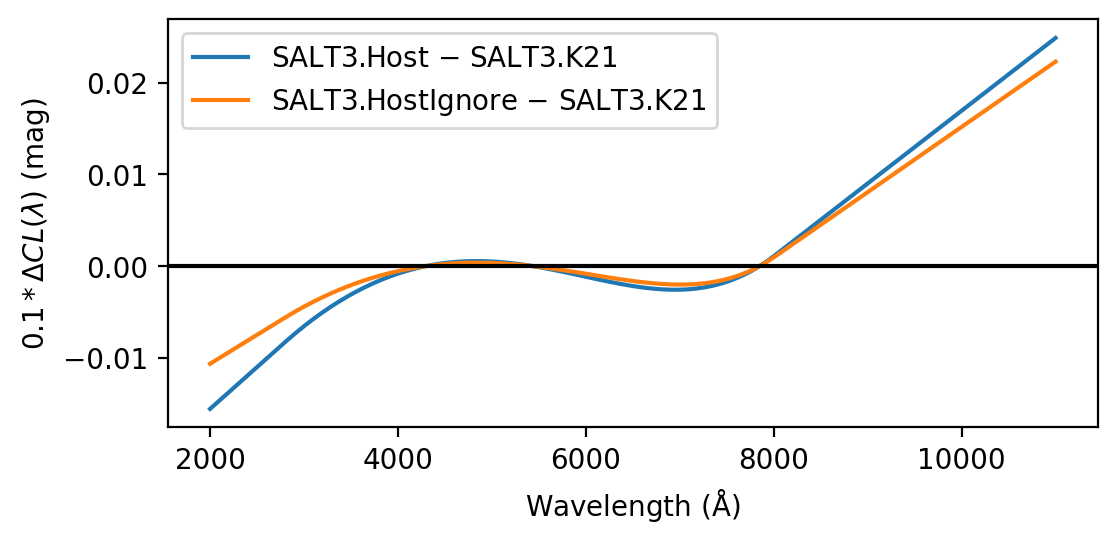}
    \caption{Difference in color laws for a $c = 0.1$ SN between K21 and the {\tt SALT3.Host} (blue) and {\tt SALT3.HostIgnore} models (orange; note that the {\tt SALT3.HostResid} model has the same color law as {\tt SALT3.HostIgnore} by construction).  {\tt SALT3.Host} and {\tt SALT3.HostIgnore} differ at the mmag level at the bluest and reddest wavelengths.}
    \label{fig:colorlaws}
\end{figure}

\begin{figure}
    \centering
    \includegraphics[width=3.5in]{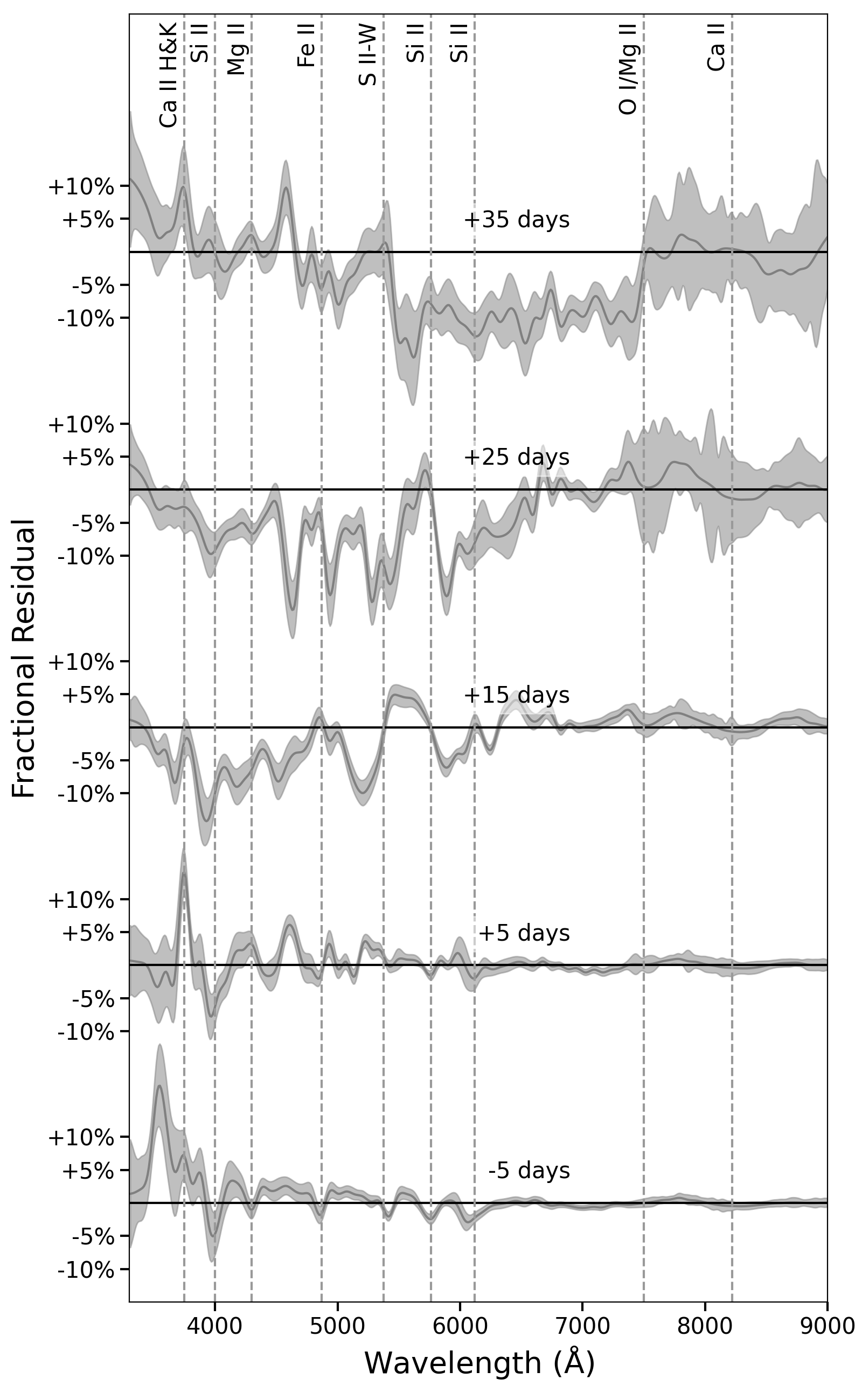}
    \caption{The evolution of the host-galaxy mass component of the {\tt SALT3.Host} spectral model with phase, which can be interpreted as  the mean spectrum of a SN\,Ia in a high-mass galaxy minus the mean spectrum of a SN\,Ia in a low-mass galaxy.  At each phase, we show $M_{host}$ as a fraction of the corresponding $M_0$ flux at 4300\AA\ (approximately the $B$-band wavelength). Variation in the line profiles, particularly those of the \ion{Ca}{H\&K} and \ion{Si}{II} features, are observed across many phases.  The tilt (color) of the spectrum also changes significantly with phase.}
    \label{fig:spec_sequence}
\end{figure}

Finally, we note that the trained color law itself is nearly identical in the {\tt SALT3.Host} model as it is in the model without a host-galaxy component (\nohost); for a red SN of $c = 0.1$, for example, the difference in color laws would amount to a change in flux of less than 0.5\% at any wavelength (Figure \ref{fig:colorlaws}).

\subsection{Model Validation}
By using the Bayesian information criterion with the log-likelihood returned by {\tt SALTshaker}, we find that the {\tt SALT3.Host} model is heavily favored over the \nohost\ model, which is trained using the original SALT formalism on the same data set.  However, we find that the likelihood of the photometric data alone changes less significantly between \nohost\ and {\tt SALT3.Host}, as discussed in Section \ref{sec:distances} below.

To ensure that host-galaxy-dependent differences are not due to systematic errors in the {\tt SALTshaker} training procedure, we use randomly assigned masses to train ``random" versions of the {\tt SALT3.Host} model, finding that only low-significance changes in spectral features are recovered.  Those results are presented in the Appendix.

\begin{figure}
    \centering
    \includegraphics[width=3.5in]{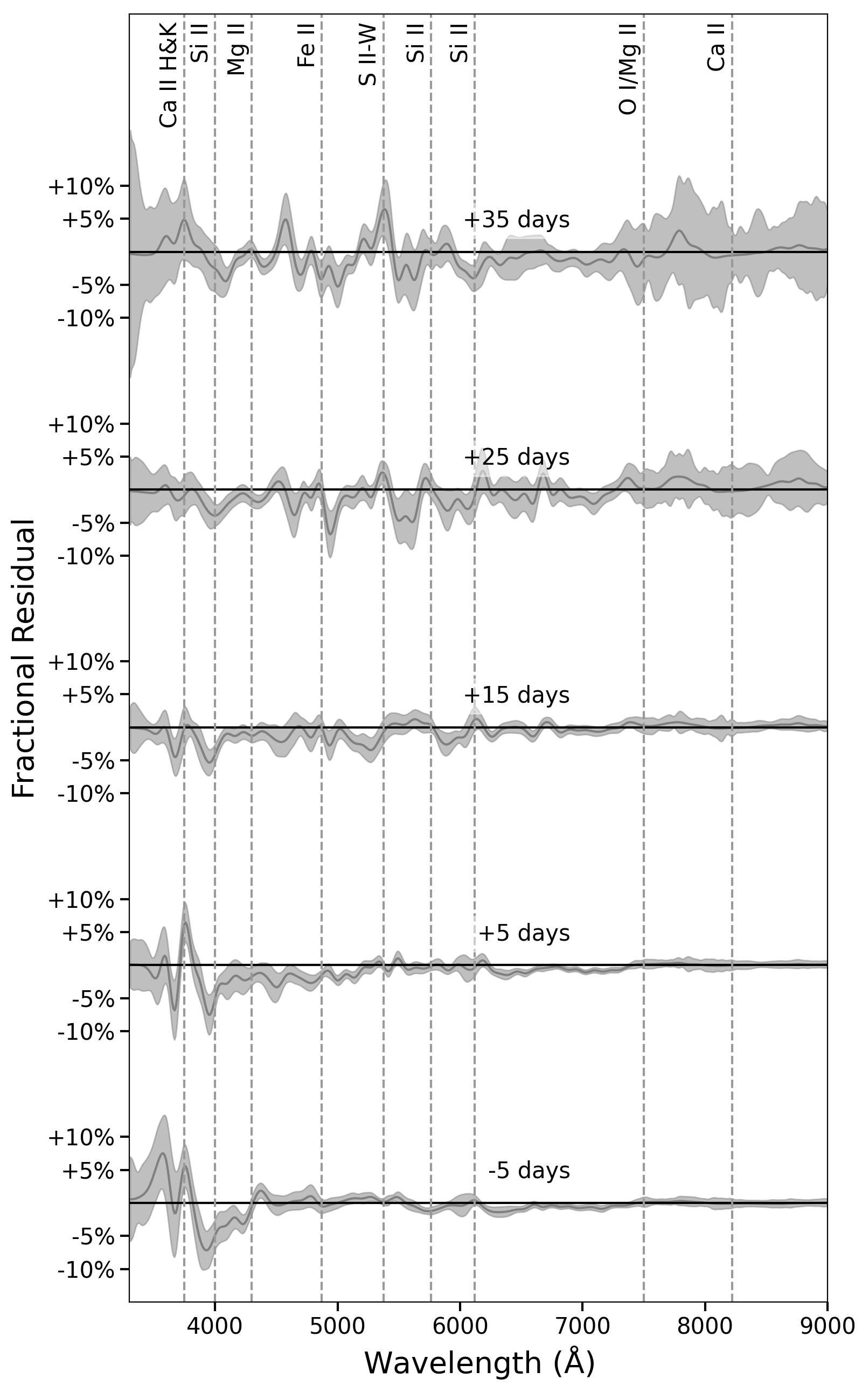}
    \caption{The same as Figure \ref{fig:spec_sequence}, with the same y-axis scale, but for the \hostnew\ model.  This model shows only the host-mass-dependent spectral component that is ``missing" from previous SALT2 and SALT3 models.  Significant spectral fluctuations are observed near maximum light in the vicinity of \ion{Ca}{H\&K}, but few statistically significant deviations are seen at redder wavelengths and later phases.}
    \label{fig:spec_sequence_fixed}
\end{figure}

\subsection{Distance Measurements}
\label{sec:distances}

Next, we use the SNANA software \citep{Kessler10} to fit the photometric training data with the {\tt SALT3.Host}, \hostnew, and \nohost\ models.\footnote{As SNANA does not yet have a host-galaxy SALT3 fitting implementation, we construct separate SALT models for SNe in high-mass and low-mass hosts from the trained {\tt SALT3.Host} and \hostnew\ models.  This is implemented by adding or subtracting half of the $M_{host}$ component from $M_0$.}
We apply selection cuts on redshift, $x_1$ and $c$ ($z > 0.01$, $-3 < x_1 < 3$, and $-0.3 < c < 0.3$, following \citealp{Scolnic18}), after which 261 of the original 296 SNe remain for measuring distances.

We subsequently estimate nuisance parameters and distances from the Tripp estimator \citep{Tripp98}:

\begin{equation}
    \mu = m_B + \alpha \cdot x_1 - \beta \cdot c -\mathcal{M}.
\end{equation}

\noindent The $\alpha$ and $\beta$ variables are nuisance parameters that standardize the SN brightness and are determined in a global fit using SNANA's {\tt SALT2mu} method \citep{Marriner11}.  $\mathcal{M}$ is the absolute magnitude of a SN\,Ia, which is set to a nominal value in this work, and $m_B$ is the log of the fitted light-curve amplitude.  The mass step, $\Delta_M$, can be included in this equation, but we treat it separately in this work.  Additionally, bias corrections to correct the measured $\mu$ for selection effects as a function of redshift are commonly included in the Tripp estimator but are not necessary here; we only examine the differential distance measurements between models and therefore this term cancels to first order.

\begin{figure}
    \centering
    \includegraphics[width=3.5in]{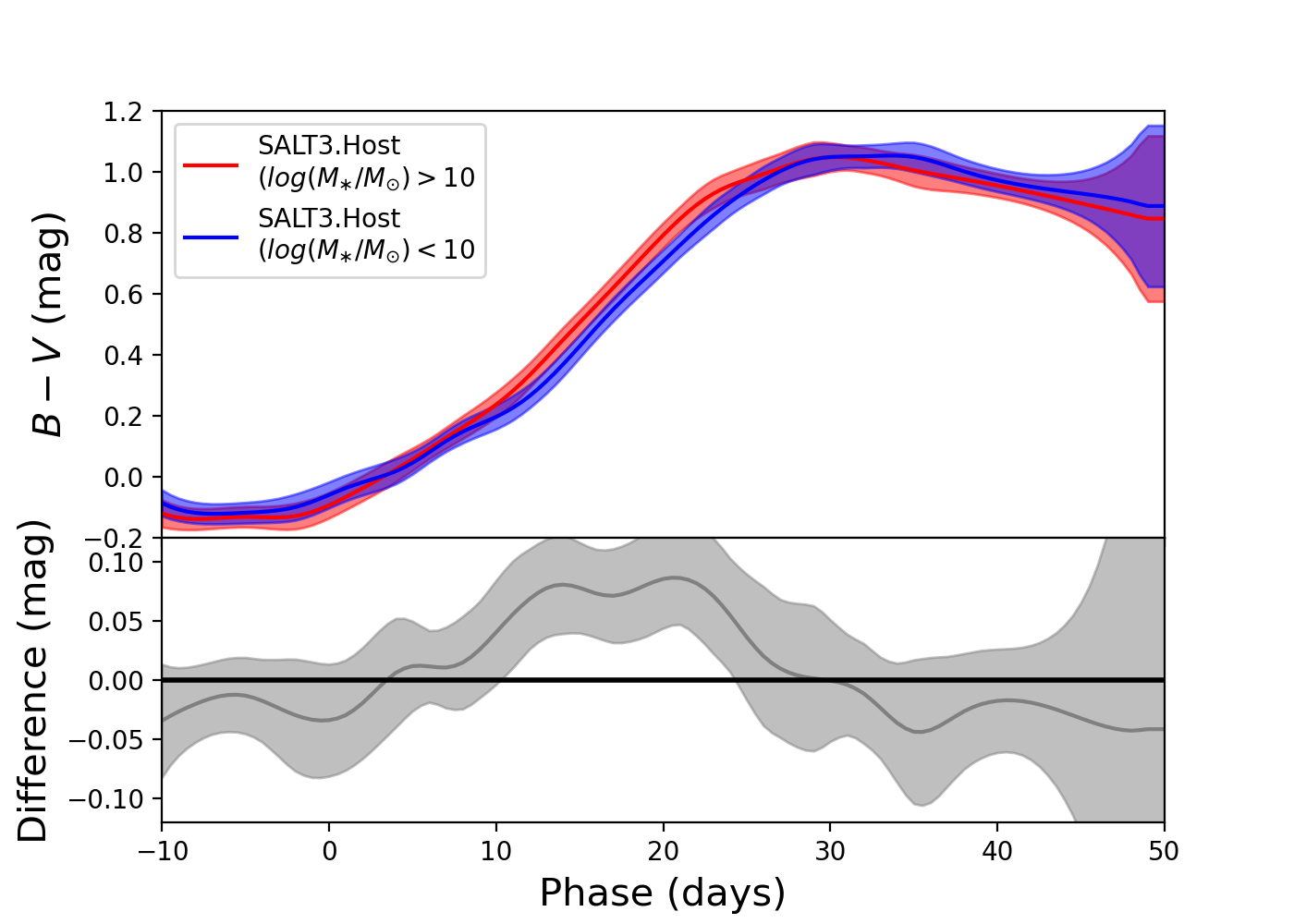}
    \caption{The $B-V$ color of the {\tt SALT3.Host} model in high- and low-mass host galaxies (top) and the difference between the two (high-mass minus low-mass colors; bottom).  There is significantly different color evolution from $\sim$10 to 20 days for SNe\,Ia in low- vs.\ high-mass hosts, which can also be seen when comparing the different light-curve shapes in Figure \ref{fig:lc}.  We note that the red and blue curves in the top panel include the $M_0$ uncertainty, while the residual curve neglects this correlated error and shows only the uncertainty on the difference (the $M_{host}$ error).}
    \label{fig:color}
\end{figure}

\begin{table*}[]
    \centering
    \caption{SALT3 Light-Curve Fitting $\chi^2$, Hubble Residual rms, and Nuisance Parameter Comparisons}
    \begin{tabular}{lrrrrr}
    \hline \hline\\*[-1.5ex]
    Version&HR rms&LC $\chi^2/\nu$~$^{\rm a}$&$\alpha$&$\beta$~$^{\rm b}$&$\Delta_M$\\
    &(mag)&&&&(mag)\\
    \hline\\*[-1.5ex]
\nohost&0.145&1.49&$0.133\pm0.009$&$2.675\pm0.092$&$0.057\pm0.018$\\
$-$ $\rm log(M_{\ast}/M_{\odot}) < 10$&0.140&1.49&$0.133\pm0.016$&$2.900\pm0.140$&\nodata\\
$-$ $\rm log(M_{\ast}/M_{\odot}) > 10$&0.147&1.49&$0.149\pm0.011$&$2.552\pm0.119$&\nodata\\
\hostnew&0.148&1.48&$0.137\pm0.009$&$2.616\pm0.093$&$0.032\pm0.019$\\
$-$ $\rm log(M_{\ast}/M_{\odot}) < 10$&0.151&1.47&$0.124\pm0.015$&$2.828\pm0.145$&\nodata\\
$-$ $\rm log(M_{\ast}/M_{\odot}) > 10$&0.147&1.50&$0.153\pm0.011$&$2.482\pm0.119$&\nodata\\
{\tt SALT3.Host}&0.144&1.45&$0.132\pm0.009$&$2.659\pm0.090$&$0.036\pm0.019$\\
$-$ $\rm log(M_{\ast}/M_{\odot}) < 10$&0.141&1.44&$0.125\pm0.020$&$2.910\pm0.134$&\nodata\\
$-$ $\rm log(M_{\ast}/M_{\odot}) > 10$&0.146&1.49&$0.137\pm0.011$&$2.549\pm0.119$&\nodata\\

\hline\\*[-1.5ex]
    \multicolumn{6}{l}{
        \begin{minipage}{6in}
        $^{\rm a}$ Neglecting model uncertainties and color dispersion terms, which are estimated such that each model has a reduced $\chi^2 \simeq 1$.  To avoid biases to the $\chi^2$, we add a 0.1\% error floor to the data uncertainties and omit $z$- and $U/u$-band data because they have significantly higher model uncertainties than the rest of the wavelength range.
        
        $^{\rm b}$ With SALT2.4 \citep{Betoule14}, we measure a consistent $\beta = 2.73 \pm 0.099$ from this sample, slightly lower than typical analyses including high-$z$ SNe ($\beta \simeq 3$; \citealp{Scolnic18}).
        \end{minipage}
        }
    \end{tabular}

    \label{table:distances}
\end{table*}

We report the measured nuisance parameters $\alpha$, $\beta$, and $\Delta_M$ in Table \ref{table:distances}.  Each parameter is computed both for the complete sample and for SNe in high-mass and low-mass host galaxies separately.  We also report the rms dispersion of Hubble residuals for each model.  We see statistically insignificant differences of just $\sim$0.01 in the $\alpha$ parameter and $\sim$0.005 in the $\beta$ parameter between {\tt SALT3.Host} and \nohost.  Between low-mass and high-mass hosts, however, we see marginally significant differences of $\sim$0.015-0.02 in $\alpha$ and $\sim$0.35 in $\beta$, suggesting that fitting $\alpha$ and $\beta$ separately for these samples might reduce distance uncertainty and perhaps cosmological bias \citep{Kelsey22}.  After subtracting the $\Lambda$CDM cosmological model prediction from \citet{Planck18}, we see a consistent Hubble residual RMS for all three models.

\begin{figure}
    \centering
    \includegraphics[width=3.35in]{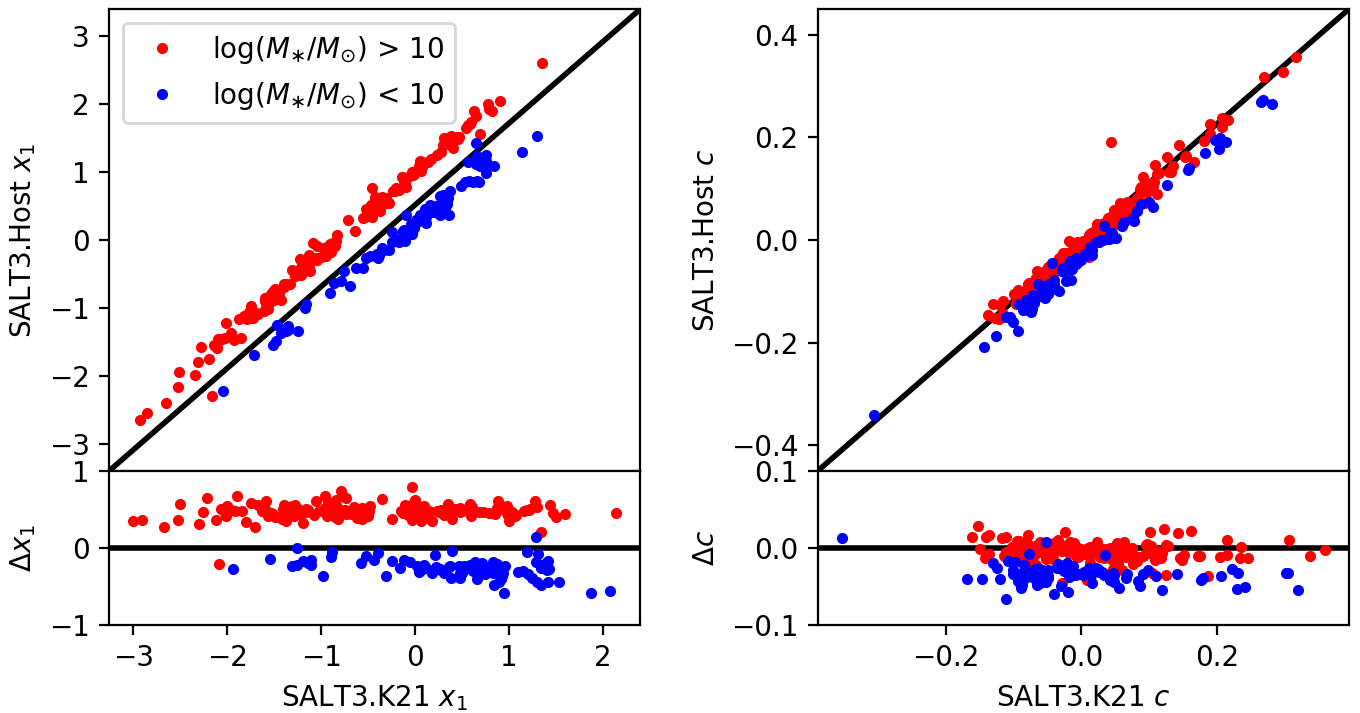}
    \caption{Comparing the $x_1$ (left) and $c$ (right) parameters measured from {\tt SALT3.Host} versus the SALT3 model from \citetalias{Kenworthy21}.  Low-mass host galaxies are shown in blue and high-mass hosts are in red.  Residuals are shown in the bottom panels.  The large difference in measured $x_1$ (left panels) is due to the removal of the $x_1$--host mass correlation present in previous SALT models.}
    \label{fig:x1ccomp}
\end{figure}

To compare {\tt SALT3.Host} to the previously published SALT3 model \citepalias{Kenworthy21}, we show the $x_1$ and $c$ parameters measured with each model in Figure \ref{fig:x1ccomp}.  The addition of a host-galaxy mass component changes the definition of $x_1$ such that there is a much lower mean difference in $x_1$ between high- and low-mass host galaxies for {\tt SALT3.Host} ($\simeq 0.17$) compared to the {\tt SALT3} model from \citetalias{Kenworthy21} ($\simeq 0.91$; hereafter {\tt SALT3.K21})\footnote{We have defined the $x_1$/$x_{host}$ correlation to be zero in {\tt SALT3.Host} training sample but --- likely due to the exact analyzed sample and treatment of uncertainties in the light-curve fitting procedure --- we find that a small residual correlation of 0.046 remains.}.  The marginally significant host-dependent variation in the $\alpha$ parameter (0.015-0.02) seems to suggest that the $x_1$--luminosity relation is changing slightly between different host types.  

Interestingly, the mean color of SNe in high-mass hosts is slightly more red ($\simeq$0.05) in {\tt SALT3.Host} compared to {\tt SALT3.K21} ($\simeq$0.03; Figure \ref{fig:x1ccomp}).  This may mean that some variation in the dust color law or the intrinsic colors of SNe that was not previously modeled may have been captured by the new $M_{host}$ component.

\begin{figure*}
    \centering
    \includegraphics[width=7in]{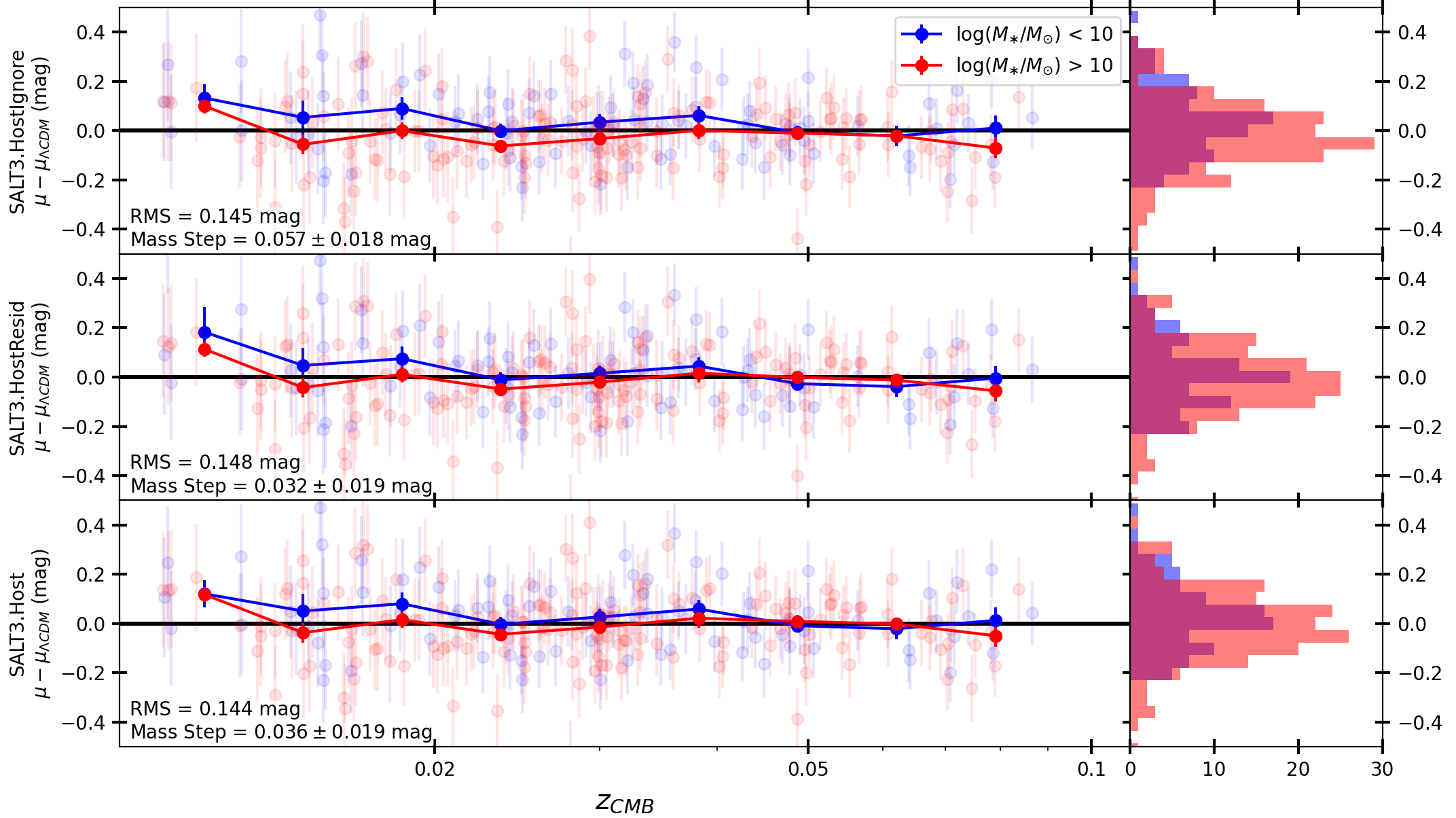}
    \caption{From top to bottom, Hubble residuals for the \nohost\ model, the \hostnew\ model, and the {\tt SALT3.Host} model.  Blue points show Hubble residuals of SNe in low-mass host galaxies, while red points show Hubble residuals of SNe in high-mass hosts.  The size of the mass step is decreased by $0.021 \pm 0.002$~mag when adding a host galaxy component to the SALT model framework.}
    \label{fig:hubbleresid}
\end{figure*}

\begin{figure}
    \centering
    \includegraphics[width=3.4in]{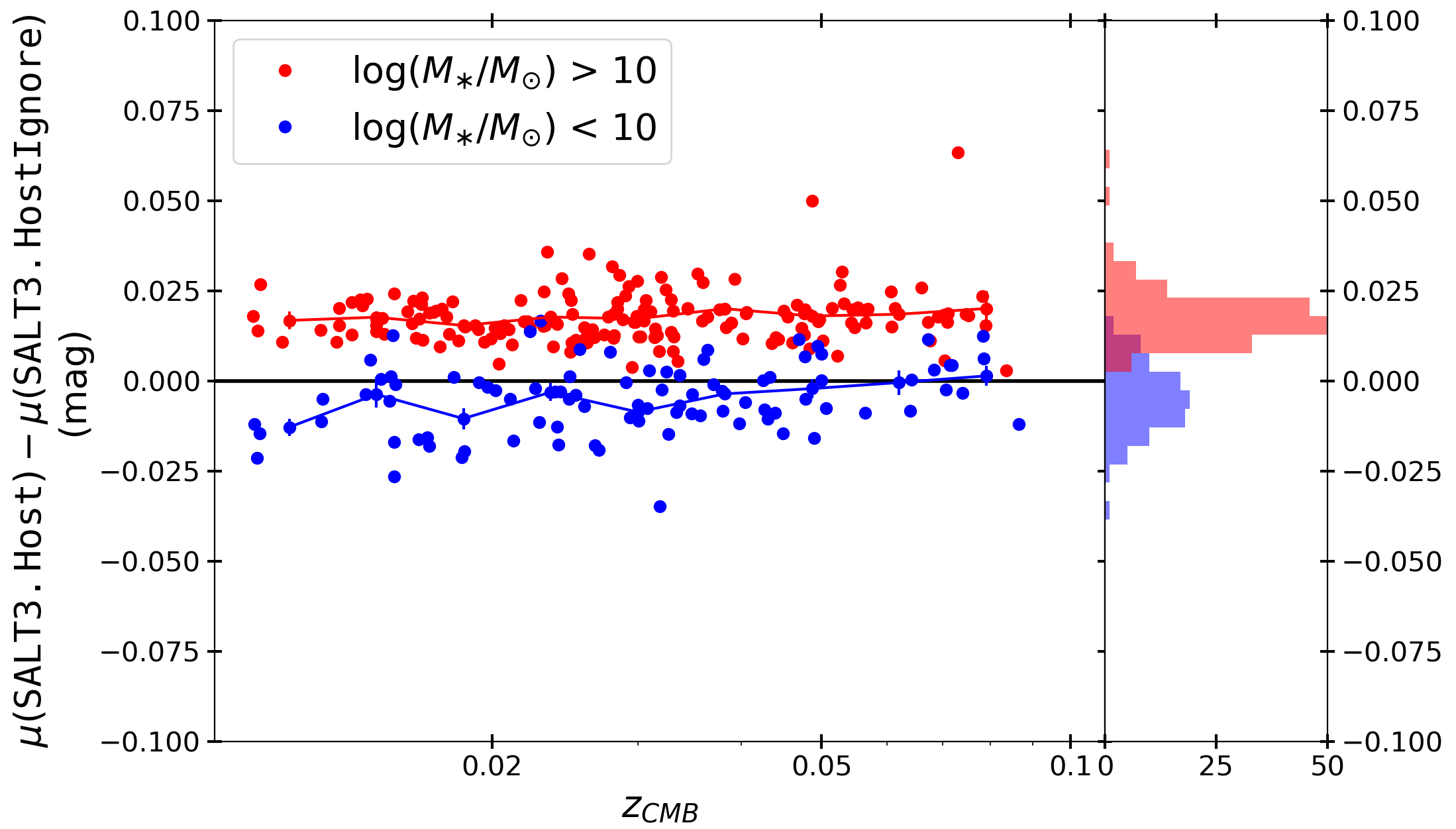}
    \caption{Difference in Hubble residuals for high-mass-hosted SNe (red) and low-mass-hosted SNe (blue) between the {\tt SALT3.Host} and \nohost\ models.  Corresponding differences in $x_1$ and $c$ parameters are shown in Figure \ref{fig:x1ccomp}.}
    \label{fig:hrdiff}
\end{figure}

\begin{figure*}
    \centering
    \includegraphics[width=7in]{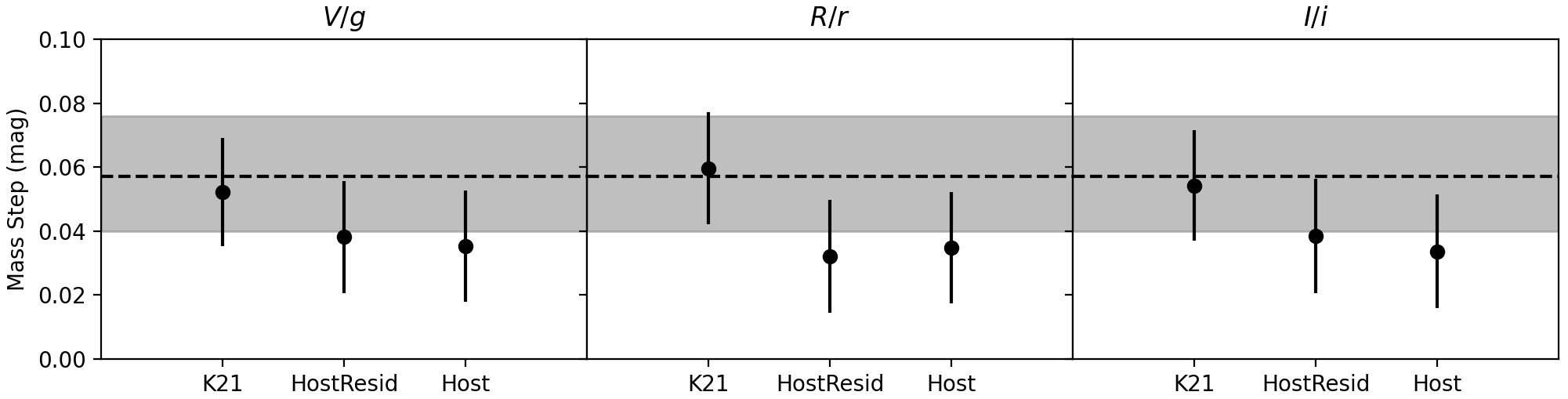}
    \caption{Mass steps measured by using $x_0$ parameters measured from the $V/g$ bands only (left), the $R/r$ bands only (center), and the $I/i$ bands only (right) for the SALT3 model from \citetalias{Kenworthy21}, the \hostnew\ model, and the {\tt SALT3.Host} model.  The gray dashed line and shaded region is the best-fit value and uncertainty from the full data set fit with the {\tt SALT3.K21} model.  Values of the $x_1$ and $c$ parameters are determined using a global fit to all available data for each SN.  Mass step measurements appear to be wavelength-independent within the measurement noise but we cannot rule out significant variation.}
    \label{fig:massstep}
\end{figure*}

\subsubsection{The Host-Galaxy Mass Step}

We measure the host-galaxy mass step by following the approach outlined in \citet{Jones18} that incorporates both Hubble residual and host-galaxy mass uncertainties. \citet{Jones18} construct a model to find the maximum-likelihood value of the step by assuming two Gaussian SN populations in Hubble residual space; the means and dispersions of these Gaussians are free parameters (see also \citealp{Rigault15} and \citealp{Jones15}).  The model takes into account that each SN has a probability of being in either a high-mass or low-mass host galaxy, which is given by the probability distribution function of the host-galaxy mass measurement.

From the \nohost\ model, we measure a mass step of $0.057 \pm 0.018$~mag versus $0.036 \pm 0.019$ mag from the {\tt SALT3.Host} model (Figure \ref{fig:hubbleresid}).  After recomputing the mass step with 100 bootstrap iterations of the SN sample, which takes into account the fact that the same data are used for both mass-step measurements, we find that the mass step is reduced by a highly significant $0.021 \pm 0.002$~mag when the {\tt SALT3.Host} model is applied.

Even with the {\tt SALT3.Host} model, the mass step remains significant at the $\sim$2-$\sigma$ level.  This implies a significant difference in the shape- and color-corrected $B$-band absolute magnitude of SNe\,Ia as a function of host mass even after explicit modeling of the SN\,Ia--host mass relation in a luminosity-independent way.  Given the low uncertainties on the mass-step difference above, we consider it likely that larger samples would measure the mass step with high statistical significance even when using the {\tt SALT3.Host} model.  

In Figure \ref{fig:hrdiff}, we show the difference in Hubble residuals between the {\tt SALT3.Host} and \nohost\ models.  Both the {\tt SALT3.Host} and \hostnew\ models change the Hubble residuals by shifting the measured color parameter $c$ (Figure \ref{fig:x1ccomp}); this appears to be because some color differences previously attributed to the global, phase-independent color are now determined by {\tt SALTshaker} to be dependent on host properties.  Additionally, the {\tt SALT3.Host} model changes the measurement of $x_1$ and therefore some $x_1$ differences are now modeled as part of the host component.  The measured $m_B$ does not change significantly in either model, which is unsurprising given that $m_B$ is defined to be the same at maximum light for both models.

These results show that our host-dependent SN\,Ia model alters the determinations of color and first-order variability, which in turn causes the mass step to shrink.
These redefined terms appear to improve the Hubble residual scatter very slightly, and we find specifically that there is less Hubble residual scatter due to the $\alpha x_1$ or $\beta c$ terms in the Tripp equation when applying the {\tt SALT3.Host} framework (a small $\sim$1-2\% improvement).  Our distance fitting and mass-step results are summarized in Figure \ref{fig:hubbleresid} and Table \ref{table:distances}.

\subsubsection{Light-Curve Fitting Residuals}

We find that the light-curve fit residuals themselves are only weakly affected by the choice in model, as shown in Table \ref{table:distances}.
When model uncertainties are completely excluded from the $\chi^2$ computation, we see that the median reduced $\chi^2$ values for the $B$, $V$ and $R$ bands are slightly reduced by $\sim$0.03-0.05.  The difference in total $\chi^2$ between models is a statistically significant $\Delta \chi^2 = 480$, but the effect on an individual SN\,Ia light curve is small.  

In addition, the median $I$ and $z$ reduced $\chi^2$ values increase slightly in the {\tt SALT3.Host} model (wavelengths at which model errors are significantly larger).\footnote{This slight increase could be an artifact of spectral recalibration uncertainties at long wavelengths, as the training code must minimize the total $\chi^2$ including both photometry and spectra.  In principle, the spectra are recalibrated as a function of wavelength, but in practice this procedure has uncertainties near the ends of the model wavelength range \citep{Dai23} and could be improved in future work.} We note that the {\it total} photometric $\chi^2$ for $z$-band photometry is slightly lower for the {\tt SALT3.Host} model, but the median reduced $\chi^2$ per SN is slightly higher, showing that some SNe have much larger contributions to the total $\chi^2$ than others.  The \hostnew\ model has a lower photometric $\chi^2$ in every band compared to \nohost, but the $\chi^2$ is larger overall than it is for {\tt SALT3.Host}.

\subsubsection{The Wavelength Dependence of the Mass Step}

Finally, we explore the wavelength dependence of the mass step by performing a global mass-step fit to the SN data using the {\tt SALT3.K21} model, the \hostnew\ model, and the {\tt SALT3.Host} model.  Next, we use the global $x_1$ and $c$ parameters measured using a fit to all bands for each SN, but generate amplitudes ($x_0$) from the $V/g$ bands, the $R/r$ bands, and the $I/i$ bands separately.  The goal is to see if the mass step changes significantly in different bands, with the different models explored here, and within the wavelengths covered by the SALT model.  The results are summarized in Figure \ref{fig:massstep}\footnote{The $z$-band result (not pictured) is consistent with results from the other bands but noisier, due to the limited $z$-band data.}.

For the K21 model, the mass step is consistent with no wavelength dependence, but we cannot rule out significant variation.  Although well within the statistical errors, the mass step is very slightly larger in the $R/r$ and $I/i$ bands than it is in the $V/g$ bands, as would be predicted from Figure \ref{fig:lc}.  Counterintuitively, a larger mass step at red wavelengths is consistent with a lower $R_V$ in high-mass hosts, as the amplitude of a SN in these redder bands is extrapolated to the $B$ band assuming the host-independent {\tt SALT3.K21} model.  We also see again that the addition of the host-galaxy mass component reduces the mass step by $\sim$0.02~mag; as the mass component is normalized to be equal to zero at $B$-band maximum light, this reduction cannot be due to luminosity information encoded in the mass step (e.g., dust extinction differences or other physical mechanisms that change the $B$-band SN luminosity).  

We also investigate the dependence of the Hubble residual rms on the color parameter $c$ to see if the trend of increased color scatter with $c$ observed by \citet{Brout21} is altered by applying the {\tt SALT3.Host} model.  We see no significant change, with both sets of distance residuals from \nohost\ and {\tt SALT3.Host} having an rms that increases from $\sim 0.13$~mag at $c \simeq -0.05$ to $\sim 0.16$~mag at $c \simeq 0.1$\footnote{Our rms is somewhat smaller than the full-sample \citet{Brout21} results, which go from $\sim 0.1$ to $\sim 0.17$~mag across this same color range.  However, the difference is likely consistent with sample-to-sample fluctuations and unrelated to the host-galaxy modeling procedure adopted here.}, indicating that adding a single host-mass component does not significantly affect the increased color scatter of red SNe.  This result is consistent with the \citet{Brout21} model, as our correction to the mean of the flux would be incapable of substantially addressing the variance caused by a large diversity of extinction laws.

The measured $\beta$ parameter between low- and high-mass hosts differs by $\sim$0.35 across all models, with high-mass-hosted SNe preferring lower values of $\beta$.  This is again consistent with the prediction of lower $R_V$ in high-mass hosts \citep{Brout21}.

\section{Discussion}
\label{sec:discussion}

It has long been known that the SALT framework is an incomplete method of standardizing SNe.  Light-curve models such as SNEMO \citep{Saunders18} demonstrated that additional principal components can improve SN\,Ia distances, and the addition of more sophisticated distance measurement tools have been shown to decrease the size of the mass step \citep{Boone21}.  Work from \citet{Rose20} and \citet{Rubin20} indicate that the optimal number of standardization parameters for SN\,Ia distance measurement is greater than the number currently included in SALT.  Furthermore, the recently developed SUGAR model \citep{Leget20}, a principal-component-based method with higher-order terms than SALT, also shows spectroscopic differences with SALT near the locations where the {\tt SALT3.Host} model differs (particularly near \ion{Ca}{H\&K} and the \ion{Si}{II} feature at 6355\AA).  Finally, given the known dependence of stretch and color on SN\,Ia host-galaxy properties, it is unsurprising that not all spectral dependencies were previously modeled by SALT.

In spite of these complexities, the mass step is typically modeled as a wavelength-independent parameter, which is perhaps part of the reason that different data sets covering different rest-frame wavelength ranges produce different mass-step measurements \citep[e.g.,][]{Brout19}.  This work provides a full spectrophotometric model of the mass step, which could be shifted to have a $B$-band magnitude of $0.036$ --- a reduction of $0.021 \pm 0.002$~mag compared to a SALT3 model without host-galaxy information --- to yield a mass step of zero.  Additionally, this shift applied to our {\tt SALTshaker} training results should cause the measured absolute magnitude difference at every phase and wavelength between a mean SN\,Ia in a high- versus low-mass host galaxy to be equal to zero.  
While much of this host-dependent variation was previously encoded by the $x_1$ parameter in previous SALT models, our \hostnew\ model shows that there are still subtle but statistically significant spectral differences (Figures \ref{fig:lc_si} and \ref{fig:spec_sequence_fixed}) and possibly subtle color differences (Figure \ref{fig:lc}, bottom panel) between high- versus low-mass-hosted SNe that were not captured by the previous SALT formalism.

\subsection{Implications for SN\,Ia Distance Measurement Methods}

Given the uncertainty in SN\,Ia physics and progenitor systems, a model that depends only on the host-galaxy mass is unlikely to be truly optimal for cosmological distance fitting.  Alternative SALT3 models that depend on specific star formation rate, progenitor age, or other properties could be used within the {\tt SALT3.Host} framework.  However, all of these models will be subject to uncertainty regarding whether they remain valid as a function of redshift as SN\,Ia progenitors and host-galaxy dust properties evolve.  A logical additional extension would be to incorporate the full host-galaxy parameter probability distribution functions as priors into the model training, allowing a model to be trained as a function of galaxy properties that have larger uncertainties.

Given that the {\tt SALT3.Host} does not yield clear differences in Hubble residuals or improvements in dispersion,  and only results in a small improvement in the quality of the light-curve fits themselves, it may be appropriate to continue using the simpler model formalism of the \nohost\ model for cosmology.  However, for measuring $w$, the leverage that the {\tt SALT3.Host} model provides may be extremely useful as a {\it systematic} test that helps us understand the wavelength ranges, redshifts, and filter sets that are particularly sensitive to the observed spectral variations as a function of host-mass properties.  As the model itself can be trained at any redshift, it is also possible to measure whether the SN\,Ia--host mass correlation is evolving with redshift as a function of both wavelength and phase.

Finally, the lack of a large difference in the goodness-of-fit statistic when comparing {\tt SALT3.Host} to \nohost\ is somewhat surprising given that both SN\,Ia spectra and light curves vary significantly as a function of their host-galaxy properties.  Given that spectra are more homogeneous in particular host-galaxy types, this result may limit the precision one could expect from a ``twinning" approach in which SNe with similar spectra are compared to yield more precise relative distance estimates \citep[e.g.,][]{Fakhouri15,Boone21}.  However, it is possible that modeling the SN spectra as a function of alternative global or local host-galaxy properties might result in more drastic improvements to Hubble residuals.

\subsection{Implications for SN\,Ia Physics}

The {\tt SALT3.Host} model gives a more precise understanding of the ways in which a mean SN\,Ia spectrum changes depending on its host galaxy.  Though the dependence of SN\,Ia properties on host-galaxy properties is already well known, by constraining the host dependence at the same time as the principal-component-like $M_0$ and $M_1$ surfaces and phase-independent color law, our model is constructed in such a way that it separates the component of SN\,Ia intrinsic diversity {\it uncorrelated} with host mass from the change in mean properties as a function of host mass.

At 1.3-$\sigma$ significance, we find evidence that the \ion{Ca}{H\&K} velocity is less blueshifted in higher-mass hosts, indicating at low significance --- and contrary to results from composite spectra in \citet{Siebert19} --- that SNe in higher-mass host galaxies may have lower explosion energies per unit mass (the difference could be due to our more robust prescription of controlling for intrinsic variation when comparing high- to low-mass-hosted SNe).  Similar evidence is seen from the larger observed equivalent widths in the Ca NIR triplet, \ion{Ca}{H\&K}, and \ion{Si}{II} line profiles, at 2.7-, 2.3-, and 2.2-$\sigma$ significance.  At late times, we note that \citet{Siebert20} found evidence for larger equivalent widths in spectral features for SNe with positive Hubble residuals (more likely to be low-mass hosts), which aligns with our results from spectra nearer to peak.  \citet{Siebert19} also saw hints of increased \ion{Ca}{H\&K} and \ion{Si}{II} absorption in late-type galaxies, which is consistent with our findings.  Finally, we see that the $I$-band secondary maximum may be slightly earlier relative to the peak flux, which would indicate a larger synthesized mass of electron-capture elements in high-mass hosts \citep{Kasen06}.  As a caveat, we note that the \ion{Ca}{H\&K} line may be blended with \ion{Si}{II} at 3858\AA\ \citep{Foley13lines}, which may have some effect on the Ca H\&K results.

Building this model as a function of other host-galaxy parameters may help to isolate additional physical processes in SN\,Ia explosions that depend on the metallicity or progenitor properties that are more prevalent in certain galaxy types.  We hope that this approach will be effective at isolating subtle SN\,Ia physical behaviors with the use of large samples.  We also hope that future investigations can continue to explore the ways in which differences in, for example, explosion energies, could affect the spectra, light curves, and luminosities of SNe in cosmological samples.

\subsubsection{Constraining the Role of Dust in the SN\,Ia Mass Step}

\citet{Brout21} predicted that the mass step is caused by reddening law variation in SN\,Ia hosts.  To generate this prediction, they modeled SN Hubble residual data assuming a symmetric intrinsic color distribution and a one-sided reddening law to break the degeneracy between intrinsic and extrinsic colors. In the {\tt SALT3.Host} model, on the other hand, we do not have the ability to conclusively separate intrinsic and extrinsic colors.  Although some substantial color differences appear to be phase dependent (Figure \ref{fig:color}), showing that intrinsic colors vary with host mass, the degree to which {\it extrinsic} colors vary is not clear from this model.  

To break this degeneracy, we would require a more sophisticated treatment of color within the SALT3 framework.  Expanding the {\tt SALT3.Host} model further by training two separate color laws for low- and high-mass host galaxies might help us to understand the degree to which a phase-independent SN color term is evolving between different galaxy types.  In addition, a SALT3 model that also includes NIR wavelengths to anchor the color, e.g., \citet{Pierel22}, will also contribute to a better understanding of the role of SN color in causing the apparent host-mass step.  

 Lastly, because the host-galaxy component of the {\tt SALT3.Host} model is shifted to have zero $B$-band flux, applying the {\tt SALT3.Host} model cannot correct for physical processes that affect the $B$-band luminosity.  Therefore, the $\sim$0.02~mag reduction in the host-galaxy mass step when using the {\tt SALT3.Host} model implies that a small but significant fraction of the mass step likely cannot be explained by reddening law differences.

\subsection{Caveats}

The training sample used here is subject to significant selection effects.  In particular, approximately half the data are comprised of samples that were chosen from a preselected set of bright galaxies; controlling for the mass step alone may not sufficiently remove the biases in this sample as a function of host-galaxy properties.  The way in which the demographics of those data could affect the recovered spectral model are difficult to fully understand from our limited data set; however, these data are the same as those used in other recent SALT model trainings \citep[e.g.,][]{Betoule14,Kenworthy21,Taylor21}, meaning that these types of biases are already present in existing models used for SN\,Ia cosmology studies.\footnote{While other recent SALT trainings also include high-redshift data, the majority of the spectroscopy originates from low-$z$ SNe.}

Despite these selection effects, spectroscopic data are important for models that explore the spectroscopic dependence of the SALT3 model on host-galaxy properties, and there exists no high-S/N sample of SN spectra comparable to the one that has been assembled over the last three decades at low redshift.  Replacing this sample with future, less biased spectroscopic data sets will take considerable work.

{\tt SALTshaker} also does not yet have a process to correct the data or model for observational biases, which could potentially be addressed by a fitting process that includes an iteratively updated, simulation-based bias-correction process to refine, for example, $x_1$ and $c$ estimations and correct the resulting model surfaces for biases in those parameters.  
Understanding the role that such biases could play in the training process are an important consideration for future work \citep{Dai23}.

Lastly, our data set remains limited redward of $\sim$7000-8000\AA.  Foundation alone has rest-frame $z$-band data to constrain, for example, the SN model in the vicinity of the \ion{Ca}{II} NIR triplet.  However, ongoing surveys such as the Young Supernova Experiment \citep{Jones21} and future surveys including the Rubin Observatory's Legacy Survey of Space and Time will supplement these data and --- with additional spectroscopic follow-up observations --- allow a better understanding of spectroscopic behavior at longer wavelengths.  Additionally, recent SN\,Ia NIR model extensions and spectroscopic data releases \citep{Lu23,Pierel22} will help to constrain the model's dependence on host-galaxy properties at $YJH$ wavelengths in the near future.

\section{Conclusions}
\label{sec:conclusions}

In this work, we expand the SALT3 framework by building a wavelength- and phase-dependent model of the SN\,Ia dependence on host-galaxy mass.  We observe variations in SN\,Ia spectral features and colors that were not previously captured by the SALT modeling framework.  We also observe a complex dependence of SN\,Ia light curves and spectra on their host-galaxy properties.  We hope that this modeling framework can be used to better understand how cosmological parameter measurements and SN selection effects may be systematically affected by the wavelength and phase dependence of the SN\,Ia---host galaxy relation.

This revised model has a significantly higher likelihood given the SN\,Ia data, though it does not appear to reduce the Hubble diagram RMS dispersion.  It does, however, reduce the size of the mass step by $0.021\pm0.002$~mag, where errors take into account the fact that the same data are used with both the {\tt SALT3.Host} and \nohost\ models. This indicates that $\sim$35\% of the mass step is due to luminosity-independent effects.  Furthermore, if the host-galaxy component is shifted such that it has a $B$-band magnitude of $+$0.036~mag at maximum light, the size of the mass step in SN\,Ia distance fitting becomes consistent with zero.  However, this value must be determined by a Hubble residual analysis and cannot be estimated by the luminosity-independent {\tt SALTshaker} model training process.

We see evidence for less energetic SN\,Ia explosions per unit mass in high-mass host galaxies, and marginal evidence for an earlier epoch of secondary maximum light that could have implications for the physics of the explosions.
We find moderate, phase-dependent differences in color between SNe in high- and low-mass hosts that --- due to their phase dependence --- are likely intrinsic.  We hope this model, and future models constrained with additional data, will offer additional insight into the physics of SN\,Ia explosions in both the nearby universe and as a function of redshift.

This revised modeling framework also directly constrains the SN\,Ia host-galaxy mass dependence in a way that was not previously possible with existing data.  It is capable of testing the redshift dependence of SN\,Ia spectra and the redshift evolution of the spectrally resolved relationships between SNe\,Ia and their hosts.  Future data sets with high-cadence, high-S/N spectral sequences will help to better constrain the dependence of SNe\,Ia on their host-galaxy masses and other physical parameters in order to better pinpoint the physics responsible for the mass step and improve the precision of distance measurements in cosmology analyses.

\begin{acknowledgements}
We would like to thank D.~Scolnic, D.~Brout, C.~Ashall and the ``value-added" journal club at the University of Hawaii's Institute for Astronomy for helpful discussions.  We would also like to thank the anonymous referee for their helpful suggestions.  D.O.J.\ is supported by NASA through Hubble Fellowship grant HF2-51462.001 awarded by the Space Telescope Science Institute (STScI), which is operated by the Association of Universities for Research in Astronomy, Inc., for NASA, under contract NAS5-26555.  The UC Santa Cruz team is supported in part by NASA grants 14-WPS14-0048, NNG16PJ34C, and NNG17PX03C, NSF grants AST-1518052 and AST-1815935, the Gordon and Betty Moore Foundation, the Heising-Simons Foundation, and by a fellowship from the David and Lucile Packard Foundation to R.J.F.
\end{acknowledgements}

\appendix
\begin{figure}
    \centering
    \includegraphics[width=3.4in]{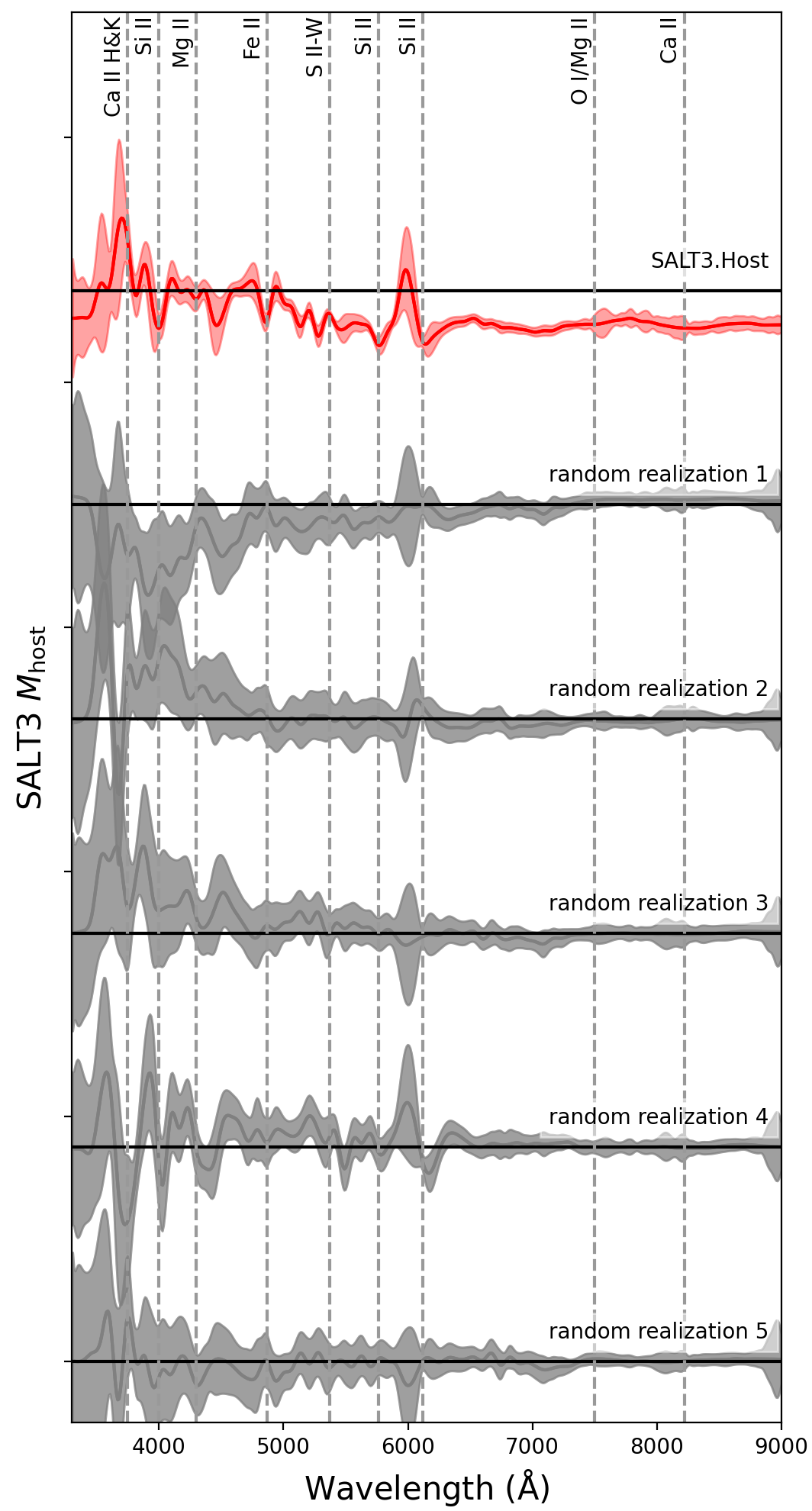}
    \caption{The trained {\tt SALT3.Host} model at maximum light (red) compared to five models trained when host-galaxy masses are assigned to each SN at random (gray).  The models using randomly assigned masses have larger errors, indicating greater variance within the 50 bootstrapping iterations, and all fluctuations are consistent with zero at the 2-$\sigma$ level.}
    \label{fig:spec_random}
\end{figure}

\begin{figure}
    \centering
    \includegraphics[width=3.4in]{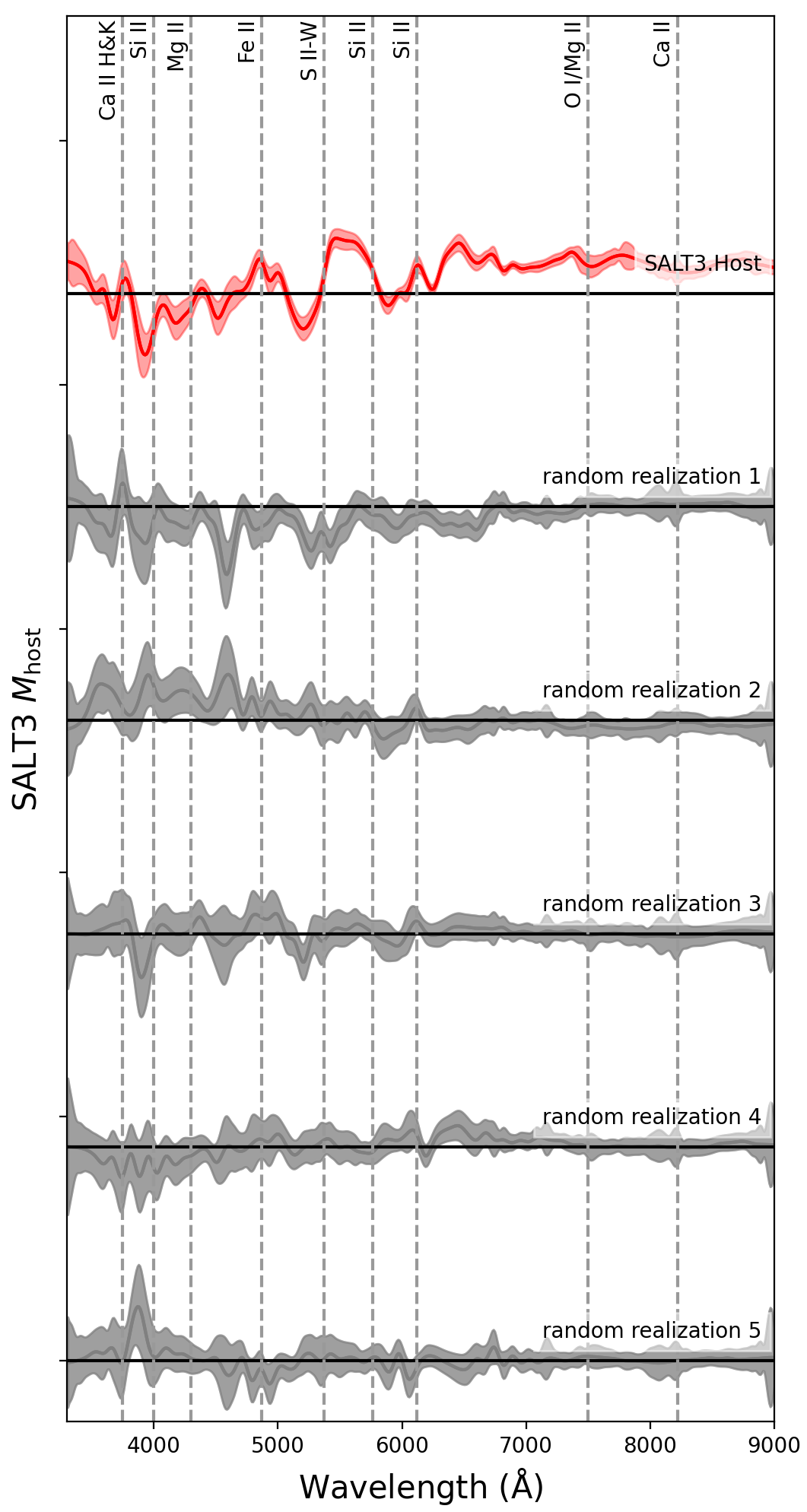}
    \caption{Same as Figure \ref{fig:spec_random} but at phase $+$15 days.}
    \label{fig:spec_random_15}
\end{figure}

To verify that the recovered, host-mass-dependent spectral features are not due to systematic errors in the {\tt SALTshaker} training, we train five {\tt SALT3.Host} models using randomly assigned host-galaxy mass parameters.  The results for maximum-light spectra are shown in Figures \ref{fig:spec_random} and \ref{fig:spec_random_15}.  We find that the errors on these random models are significantly larger, resulting from greater variance in the 50 bootstrap iterations for each ``random" model.  We do not recover any spectral features at more than 2-$\sigma$ significance, including at both earlier and later light-curve phases.  We do see hints that that some low-significance absorption features may be correlated, e.g., \ion{Si}{II} and \ion{Ca}{H\&K} in random realization 4 at maximum light, due to their correlation in the individual input spectra.  There could also be slight hints of wavelength-dependent slopes in some realizations but they are much less significant than the slope seen in the {\tt SALT3.Host} model.

We also measure the difference in equivalent widths between mean high- and low-mass-hosted SN spectra of the \ion{Ca}{H\&K} and \ion{Si}{II} (6355\AA) features for each of the five random realizations.  Out of 10 total measurements, four have $>$1-$\sigma$ significance, and one of those four is significant at $>$2$\sigma$.  This is consistent with the statistical expectation; for 10 samples in which the errors are due to random noise, one would expect an average of three 1-$\sigma$ outliers and 0.5 2-$\sigma$ outliers.

\bibliography{main}

\end{document}